\documentclass[12pt]{article}
\usepackage{amssymb,amsmath,latexsym}
\title{Phase-averaged transport\\
for quasi-periodic Hamiltonians 
\vspace{.5cm}}

\author{Jean Bellissard$^{1,2}$, Italo Guarneri$^{3,4,5}$, Hermann 
Schulz-Baldes$^{6}$\\
{\small $^1$ Universit\'e Paul-Sabatier, 118 route de Narbonne, 31062
Toulouse, France}\\
{\small $^2$ Institut Universitaire de France}\\
{\small $^3$ Universit\`a dell'Insubria a Como, via Valleggio 11, 22100 Como, 
Italy } \\
{\small $^4$ Istituto Nazionale per la Fisica della Materia,  
via Celoria 16, 20133 Milano, Italy}\\
{\small $^5$ Istituto Nazionale di Fisica Nucleare, Sezione di Pavia, 
via Bassi 6, 27100 Pavia, Italy}\\
{\small $^6$ University of California at Irvine, CA, 92697, USA}
\vspace{.2cm}}

\date{ }

\newtheorem{theo}{Theorem}
\newtheorem{defini}{Definition}
\newtheorem{proposi}{Proposition}
\newtheorem{lemma}{Lemma}

\newtheorem{rem}{Remark}

\newcommand{\Aa}{{\mathcal A}}
\newcommand{\Bb}{{\mathcal B}}
\newcommand{\Cc}{{\mathcal C}}

\newcommand{\Ff}{{\mathcal F}}
\newcommand{\Gg}{{\mathcal G}}
\newcommand{\Hh}{{\mathcal H}}

\newcommand{\Ll}{{\mathcal L}}
\newcommand{\Mm}{{\mathcal M}}
\newcommand{\Nn}{{\mathcal N}}
\newcommand{\Oo}{{\mathcal O}}

\newcommand{\Ss}{{\mathcal S}}
\newcommand{\Uu}{{\mathcal U}}
\newcommand{\Ww}{{\mathcal W}}

\newcommand{\ZZ}{{\bf Z}}

\newcommand{\lL}{{\mathbf l}}
\newcommand{\mM}{{\mathbf m}}
\newcommand{\kaK}{{\mathbf \kappa}}

\newcommand{\CM}{{\mathbb C}}
\newcommand{\NM}{{\mathbb N}}
\newcommand{\QM}{{\mathbb Q}}
\newcommand{\RM}{{\mathbb R}}

\newcommand{\TM}{{\mathbb T}}
\newcommand{\ZM}{{\mathbb Z}}

\newcommand{\aG}{{\mathfrak a}}
\newcommand{\bG}{{\mathfrak b}}

\newcommand{\mG}{{\mathfrak m}}
\newcommand{\lG}{{\mathfrak l}}

\newcommand{\WG}{{\mathfrak W}}
\newcommand{\Sssmall}{{\mathfrak s}}

\newcommand{\naV}{{\vec{\nabla}}}

\newcommand{\hT}{{\widehat{T}}}
\newcommand{\hX}{{\widehat{X}}}
\newcommand{\hU}{{\widehat{U}}}

\newcommand{\tr}{\mbox{\rm Tr}}                   
\newcommand{\TV}{{\mathcal T}}                    
\newcommand{\Ath}{\Aa_{\theta}}                   
\newcommand{\Athp}{\Aa_{\theta'}}                 
\newcommand{\Cth}{\Cc^{\infty}(\Aa_{\theta})}     
\newcommand{\Cs}{$C^{\ast}$-algebra }             
\newcommand{\1}{{\bf 1}}                          
\newcommand{\Wth}{W_{\theta}}                     
\newcommand{\Wwth}{\Ww_{\theta}}                  
\newcommand{\Wthp}{W_{\theta'}}                   
\newcommand{\Wwthp}{\Ww_{\theta'}}                
\newcommand{\piw}{\pi_W}                          
\newcommand{\fio}{\phi_0\,}                       
\newcommand{\fis}{\phi_S\,}                       
\newcommand{\rhs}{\rho_S}                         
\newcommand{\HGS}{{\mathfrak H}_{S}\,}            
\newcommand{\etah}{\widehat{\eta}}                
\newcommand{\Var}{\mbox{\rm Var}}                 
\newcommand{\Dpsi}{F^\theta_{\psi}}               
\newcommand{\Tpsi}{T^\theta_{\psi}}               
\newcommand{\Dphi}{F^\theta_{\phi}}               
\newcommand{\Tphi}{T^\theta_{\phi}}               
\newcommand{\Tphio}{T^\theta_{\phi_0}}            
\newcommand{\twod}{{\mbox{\rm\tiny 2D}}}          
\newcommand{\oned}{{\mbox{\rm\tiny 1D}}}          

\begin{document}

\maketitle

\begin{abstract}
For a class of discrete quasi-periodic Schr\"odinger operators 
defined by covariant re- presentations of the rotation algebra,  
a lower bound on phase-averaged transport in terms of the 
multifractal dimensions of the density of states is proven. This result 
is established under a Diophantine condition on the incommensuration 
parameter. The relevant class of operators is distinguished by invariance 
with respect to symmetry automorphisms of the rotation algebra. It 
includes the critical Harper (almost-Mathieu) operator. As a
by-product, a new solution of the frame problem associated with 
Weyl-Heisenberg-Gabor lattices of coherent states is given.

\end{abstract}



\vspace{1cm}

\section{Introduction}

This work is devoted to proving a lower bound on the diffusion
exponents of a class of quasiperiodic Hamiltonians in terms of the 
multifractal dimensions of 
their density of states (DOS). The class of models involved describes the 
motion of a charged particle in a perfect two-dimensional crystal with 
$3$-fold, $4$-fold or $6$-fold symmetry, submitted to a uniform irrational 
magnetic field. Irrationality means that the magnetic flux through each lattice
cell is equal to an irrational number $\theta$ in units of the flux quantum. 
As shown by Harper~\cite{Har} in the specific case of a square lattice with 
nearest neighbor hopping, the Landau gauge allows to reduce such models 
to a family of Hamiltonians each describing the 
motion of a particle on a $1D$ chain with quasiperiodic 
potential. The latter representation gives a 
strongly continuous family $H= (H_{\omega})_{\omega \in \TM}$ 
of self-adjoint bounded operators on the Hilbert space $\ell^2(\ZM)$
of the chain indexed by a phase $\omega\in\TM=\RM/2
\pi\ZM$. This family satisfies the covariance relation
$\hT H_{\omega}\hT^{-1}=H_{\omega + 2\pi \theta}$
(here $\hT$ represents the 
operator of translation by one site along the chain).

\vspace{.2cm}
 
The phase-averaged diffusion exponents $\beta (q)$, $q>0$, of $H$ are defined by:

$$
\int_{\TM}\, d\omega \;
 \int_{-T}^T\frac{dt}{2T}\;
  \langle \phi |
   e^{\imath H_\omega t}|\hX|^{q} e^{-\imath H_\omega t}
    |\phi \rangle 
     \;\stackrel{{\textstyle \sim}}{{\scriptscriptstyle T\uparrow\infty}}\;
      T^{q\beta (q)}
\mbox{ , } 
$$

\noindent where $\hX$ denotes the position operator on the chain.
The DOS of the family $H$ is the Borel measure 
$\Nn$ defined by phase-averaging the spectral measure with respect to any
site. Its generalized multifractal 
dimensions  $D_{\Nn}
(q)$ for $ q\neq 1$ are formally defined by

$$\int_{\RM} \,d \Nn (E)\,
   \left(
     \int_{E-\varepsilon}^{E+\varepsilon}\,
      d\Nn(E') 
   \right)^{q-1}\;
    \stackrel{{\textstyle \sim}}{{\scriptscriptstyle \varepsilon\downarrow 0}}
     \; \varepsilon^{(q-1)D_{\Nn} (q)}
\mbox{ . }
$$

\noindent A somewhat imprecise statement of 
the main result of this work is: whenever $\theta/2\pi$ is a 
Roth number \cite{Her} (namely, for any $\epsilon >0$, there is $c>0$ such 
that $|\theta - p/q|\geq c/q^{2+\epsilon}$ for all $p/q \in \QM$), and 
for the class of models mentioned above, the following inequality holds
for all $0< q <1$
\begin{equation}
\label{exponentsinequality}
\beta(q) \, \geq \, D_{\Nn} (1-q)
\mbox{ . }
\end{equation}
\noindent This result can be reformulated in terms of two-dimensional
magnetic operators on the lattice and then gives an improvement of the general
Guarneri-Combes-Last lower bound \cite{Gua,Com,Las2} by a factor 2.
More precise definitions and statements will be given in
Section~\ref{sec-notresults}.

\vspace{.2cm}

The inequality (\ref{exponentsinequality}) has been motivated by 
 work by Pi\'echon~\cite{Pie}, who gave heuristic arguments and 
numerical support for $\beta(q)=D_\Nn(1-q)$ for $q>0$, 
valid for the Harper model and the Fibonacci chain (for the latter
case, a perturbative argument was also given).
It was theoretically and numerically demonstrated by Mantica~\cite{Man}
that the same {\em exact} relation between spectral and transport
exponents is also valid for  the
Jacobi matrices associated with a Julia set. This result was 
rigorously proven in \cite{GSB1,BSB}. For the latter operators, the
DOS and the local density of states
(LDOS) coincide. 

\vspace{.2cm}

Numerous works \cite{Gua,Com,Las2,GSB2,GSB3,BGT} yield 
lower bounds on the quantum diffusion of a given wave packet in terms
of the fractal properties of the corresponding 
LDOS. These rigorous lower bounds are typically not  
optimal as shown by numerical simulations \cite{GM,Ketz}.
Better lower bounds are obtained if the behaviour of generalized 
eigenfunctions is taken into account \cite{Ketz}. Kiselev and Last 
have proven general rigorous bounds in terms of upper bounds for 
the algebraic decay of the eigenfunctions  \cite{KL}.

\vspace{.2cm}

However,  in most models used in solid state physics, the 
Hamiltonian is a covariant strongly continuous family of self-adjoint
operators \cite{Bel} indexed by a variable which represents the 
phase or the configuration of disorder. The measure class of the singular 
part 
of the LDOS may sensitively depend on the phase \cite{DS}. In addition, 
the multifractal dimensions are not even measure class invariants
\cite{SBB} (unlike the Hausdorff and packing dimensions). 
 This raises concerns about the practical relevance of  bounds based 
on multifractal dimensions of the LDOS in this context. 
The bound (\ref{exponentsinequality}) 
has a threefold advantage: (i) it involves the  DOS, which  
is phase-averaged; (ii) it does not require information about eigenfunctions;
(iii) the exponent of phase-averaged transport is the one that 
determines  the 
low temperature behaviour of the conductivity \cite{SBB}.

\vspace{.2cm}

The present formulation uses   the C$^\ast$-algebraic
framework introduced by one of the authors for the study of homogeneous
models of solid state physics. While referring to \cite{Bel,SBB} for
motivations and details, in the opening Section \ref{sec-notresults}
we briefly recall some of the basic notions. A precise statement of our 
main results is also 
given in Section \ref{sec-notresults}, along with an outline 
of the logical structure of their proofs. In the subsequent sections 
we present more results and proofs. 

\vspace{.2cm}

\noindent {\bf Acknowledgements:} We would like to thank B. Simon, R.
Seiler and S. Jitormiskaya for very useful comments. The work 
of H. S.-B. was
supported by NSF Grant DMS-0070755 and the DFG Grant SCHU 1358/1-1. 
J.B. wants to
thank the Institut Universitaire de France and the MSRI at Berkeley for
providing support while this work was in progress.

\vspace{.5cm}

\section{Notations and results}
\label{sec-notresults}

A number $\alpha\in \RM$ is of Roth type if and only if, 
for any $\epsilon>0$, there is a constant $c_\epsilon >0$ such that for all
rational numbers $p/q$ the following inequality holds

\begin{equation}
\label{eq-roth} 
 \left|\alpha-\frac{p}{q}\right| \;\geq\; 
  \frac{c_\epsilon}{q^{2+\epsilon}}\;.
\end{equation}
\noindent Most properties of numbers of Roth type can be found in \cite{Her}. They
form a set of full Lebesgue measure containing all algebraic
numbers (Roth's theorem). 
$\theta>0$ will be called a Roth angle if $\theta/2\pi$ is a 
number of Roth type.

\vspace{.2cm}

The {\em rotation algebra} $\Ath$ \cite{Rie} is the smallest
\Cs generated by two unitaries $U$ and $V$, such that $UV = e^{\imath\,
\theta}\, VU$. 
It is convenient to set  $\Wth (\mM) = e^{-\imath\, \theta\,
m_1 m_2/2} \, U^{m_1}\,V^{m_2}$, whenever $\mM = (m_1,m_2) \in \ZM^2$.
The $\Wth (\mM)$'s are unitary operators satisfying 
$\Wth
(\lL) \Wth(\mM) = e^{\imath\,(\theta/2)\, \lL\wedge\mM}\, \Wth (\lL+\mM)$
where $\lL\wedge\mM=l_1m_2-l_2m_1$. The unique trace on 
$\Ath$ ($\theta/2\pi$ irrational) is defined by 
$\TV_{\theta}(\Wth (\mM))=\delta_{\mM,0}$.
A strongly continuous action of the torus ${\TM}^2$ on $\Ath$ is given by
$((k_1,k_2),\Wth(\mM))\in {\TM^2}\times\Ath\mapsto e^{\imath(m_1k_1+m_2k_2)}
\Wth(\mM)$. The associated $\ast$-derivations are denoted by
$\delta_1,\delta_2$. For $n\in\NM$, one says $A\in\Cc^n(\Ath)$ if
$\delta_1^{m_1} \delta_2^{m_2} A\in \Ath$ for all positive integers
$m_1,m_2$ satisfying $m_1+m_2\leq n$.

\vspace{.2cm}

$\Ath$ admits three classes of representations that will be considered in this
work. The {\it $1D$-covariant representations} is a faithful family
$(\pi_{\omega} 
)_{\omega \in \RM}$ of representations on $\ell^2(\ZM)$ defined
by $\pi_{\omega}(U) =\hT$ and 
$\pi_{\omega}(V) =e^{\imath (\omega -\theta \hX)}$ where $\hT$ and $\hX$ 
are the shift and the position operator respectively, namely
$$\hT\,u(n) \;=\; u(n-1)\,,
\hspace{1cm}
   \hX\,u (n) \;=\; n\,u(n)\,,
   \hspace{2cm}
    \forall\;\; u\in\ell^2(\ZM)\,.
$$ 

\noindent It follows that $\pi_{\omega+2\pi} =\pi_{\omega}$ ({\em periodicity})
and that $\hT\pi_{\omega} (\cdot )\hT^{-1}= \pi_{\omega+\theta}(\cdot )$ ({\em
covariance}). Moreover $\omega \mapsto \pi_{\omega}(\cdot )$ is strongly
continuous. In the sequel, it will be useful to denote by $|n\rangle = u_n\,
\; (n\in\ZM)$ the canonical basis of $\ell^2 (\ZM)$ defined by $u_n(n')=
\delta_{n,n'}$. The {\it $2D$-representation} (or the GNS-representation of
$\TV_{\theta}$) is given by the magnetic translations on
$\ell^2(\ZM^2)$ (in symmetric gauge):
$$ 
\pi_\twod(\Wth (\mM))\psi(\lL)
\;=\;
e^{\imath\theta \mM\wedge \lL/2}\psi(\lL-\mM)
\mbox{ , }
\qquad
\psi\in\ell^2(\ZM^2)
\mbox{ . }
$$

\noindent The position operators
on $\ell^2(\ZM^2)$ are denoted by $(X_1,X_2)$.
The {\it Weyl representation} $\piw$ acts on $L^2(\RM)$. Let $Q$ and $P$
denote the position and momentum operators defined by $Q\phi (x) = x\phi (x)$
and $P\phi = -\imath d\phi/dx $ whenever $\phi$ belongs to the Schwartz space
$\Ss (\RM)$. It is known that $Q$ and $P$ are essentially selfadjoint and
satisfy the canonical commutation rule $[Q,P]=\imath \1$. Then $\piw$ is defined
by
$$\piw (U) \;=\; e^{\imath \sqrt{\theta}P}\,,
\hspace{2cm}
   \piw (V) \;=\; e^{\imath \sqrt{\theta}Q}\,.
$$

\noindent For every $\theta>0$, $\piw$ and $\pi_\twod$ are unitarily equivalent 
and faithful. More results  about $\Ath$ are reviewed in 
Section~\ref{ssec-rotationalg}.

\vspace{.2cm}

The group $SL(2,\ZM)$ acts on $\Ath$ through the automorphisms
$\etah_S(\Wth (\mM))=\Wth (S\mM)$, $S\in SL(2,\ZM)$. 
$S$ is called a {\em symmetry} if $S\neq \pm\1$ and
$\sup_{n\in\NM}\|S^n\| <\infty$. Of special interest
are the $3$-fold, $4$-fold and $6$-fold symmetries 
$$S_3\;=\; 
      \left(
       \begin{array}{cc}
        0 & -1 \\
        1 & -1
       \end{array}
      \right)\,,
    \hspace{1cm}
  S_4\;=\;
   \left(
    \begin{array}{cc}
     0 & -1 \\
     1 & 0
    \end{array}
   \right)\,,
    \hspace{1cm}
     S_6\;=\;
      \left(
       \begin{array}{cc}
        1 & -1 \\
        1 & 0
       \end{array}
      \right)\,,    
$$

\noindent respectively generating the symmetry groups of the hexagonal
(or honeycomb), square
and triangular lattices in  dimension $2$. 

In this work, the Hamiltonian $H=H^{\ast}$ is an element of $\Ath$. 
Of particular interest are Hamiltonians invariant under 
some symmetry $S\in SL(2,\ZM)$, that is $\etah_{S} (H) =H$.
The most prominent among such operators is 
the (critical) Harper Hamiltonian on a square lattice 
$H_4 = U+U^{-1}+V+V^{-1}$. For the sake of concreteness, let us write 
out its
covariant representations

$$
\pi_\omega(H_4)u(n)
\;=\;
u(n+1)+u(n-1)+2\cos(n\theta+\omega)u(n)
\mbox{ , }
\qquad
u\in \ell^2(\ZM)
\mbox{ . }
$$

\noindent Its Weyl representation is
$\piw(H_4)=2\,\cos(\sqrt{\theta}\,Q)+2\,
\cos(\sqrt{\theta}\,P)$.
Further examples are the magnetic operator on 
a triangular lattice $H_6 = U+U^{-1}+V+V^{-1}+
e^{-\imath\theta/2} UV + e^{-\imath\theta/2} U^{-1}V^{-1} $ 
as well as on a hexagonal
lattice (which reduces to two triangular ones \cite{Ram}). 

\vspace{.2cm}

For $H=H^*\in\Ath$ let us introduce the notations $H_\omega=\pi_\omega(H)$ and
$H_\twod=\pi_\twod(H)$. Its  {\em density of
states} (DOS) is the measure $\Nn$ defined by (see, {\sl e.g.}, \cite{Bel})

\begin{equation}
\label{eq-DOSdef}
\int_{\RM} \;
   d\Nn (E)\, f(E) \;=\;
    \TV_{\theta}\left( f(H) \right)\,
\;=\; \langle {\bf 0}|\,f(H_\twod)\,|{\bf 0}\rangle
\;=\;\lim_{\Lambda\to\infty}\frac{1}{\Lambda}\mbox{Tr}_\Lambda(H_\omega)\;,
\qquad f\in \Cc_0(\RM)\;.
\end{equation}

\noindent Here $|{\bf 0}\rangle$ denotes the normalized state localized at the origin
of $\ZM^2$, $\mbox{Tr}_\Lambda(A)=\sum_{n=1}^\Lambda 
\langle n|A|n\rangle$ and the last equality in (\ref{eq-DOSdef}) holds almost surely.
For a Borel set $\Delta\subset \RM$ and a Borel measure
$\mu$, the
family of {\em generalized multifractal dimensions} is defined by 
\begin{equation}
\label{eq-fracdimdos} 
D^\pm_\mu (\Delta; q)\;=\;
 \frac{1}{1-q} \;
  {\lim_{T\to \infty}}^{\!\!\pm} \;
   \frac{\log\left( \int_{\Delta} d\mu (E)\,
    \left(\int_\Delta d \mu(E')\, 
     \exp (-(E-E')^2 T^2)\right)^{q-1}\right)}{\log(T)}\,,
\end{equation}

\noindent where $\lim^{+}$ and 
$\lim^{-}$ denote $\limsup$ or $\liminf$ respectively.
The gaussian $\exp (-(E-E')^2 T^2)$
may be replaced by the indicator function on $[E-\frac 1T,
E+\frac 1T]$ without changing the values of the generalized dimensions
\cite{GSB3,BGT}.

\vspace{.2cm}

Let now $H\in \Cc^2(\Ath)$. The diffusion exponents of $H_\twod$
are defined by 

\begin{equation}
\label{eq-exponent2D} 
\beta_\twod^{\pm} (H,\Delta;q) \;=\;
   {\lim_{T\to \infty}}^{\!\!\pm}\; 
    \frac{\log(\langle M_\twod(H,\Delta;q,\cdot)\rangle_T)}{q \log(T)}
\mbox{ , }
\qquad
q\in(0,2]
\mbox{ , }
\end{equation}
\noindent where
\begin{equation}
\label{eq-moment2D} 
M_\twod(H,\Delta;q,t) \;=\;
   \langle {\bf 0} |\chi_{\Delta}(H_\twod) \,
     e^{\imath H_\twod t}(|X_1|^{q}+|X_2|^q) e^{-\imath H_\twod t}
      \chi_{\Delta}(H_\twod) |{\bf 0} \rangle \,,
\end{equation}

\noindent and $\langle f(\cdot)\rangle_T$ 
denotes the average $\int_{-T}^{+T} dt \,f(t)/2T$
of a measurable function $t\in\RM\mapsto f(t)\in\RM $. The
{\em phase-averaged diffusion exponents} of the covariant family
$(H_\omega)_{\omega\in\Omega}$ 
are defined as in (\ref{eq-exponent2D}) as growth exponents of

\begin{equation}
\label{eq-moment1D} 
M_\oned(H,\Delta;q,t) \;=\;\int_{\TM}\frac{d\omega}{2\pi}\;
   \langle  0 |\chi_{\Delta}(H_\omega) \,
     e^{\imath H_\omega t}|\hat{X}|^{q} e^{-\imath H_\omega t}
      \chi_{\Delta}(H_\omega) |0 \rangle \,,
\end{equation}

\noindent Because $H\in \Cc^2(\Ath)$ and $q\in (0,2]$, 
$M_\twod (H,\Delta;q,t)$ and $M_\oned (H,\Delta;q,t)$ are finite.
Moreover, $\beta_\twod^{\pm} (H,\Delta;q)$
and $\beta_\oned^{\pm} (H,\Delta;q)$ take values in the interval $[0,1]$ \cite{SBB}.

\vspace{.4cm}

\noindent {\bf Main Theorem~} {\em 
Let $\theta$ be a Roth angle and $H=H^{\ast}\in\Cc^2(\Ath)$.

\vspace{.1cm}

\noindent {\bf {\rm (i)}}
For any Borel subset $\Delta \subset \RM $ and $q\in (0,1)$
\begin{equation}
\label{eq-main1}
\beta_\twod^{\pm} (H,\Delta;q) \; \geq \;
   D_{\Nn}^{\pm}(\Delta; 1-q)\,.
\end{equation}

\noindent {\bf {\rm (ii)}}
Let $H$ be invariant under some symmetry $S\in SL(2,\ZM)$.  
Then, for any Borel subset $\Delta \subset \RM $ and $q\in (0,1)$
\begin{equation}
\label{eq-main2}
\beta_\oned^{\pm} (H,\Delta;q) \; \geq \;
   D_{\Nn}^{\pm}(\Delta; 1-q)\,.
\end{equation}

}

\vspace{.2cm}

\begin{rem}
{\rm Existing lower bounds (inequalities proved in \cite{GSB3,BGT}) yield 
$\beta_\twod^{\pm} (H,\Delta;q)\geq \frac{1}{2}D_{\Nn}^{\pm}(\Delta; 1/(1+q))
$
where the factor $\frac{1}{2}$ stems from the dimension of physical
space. In addition, $D_{\Nn}^{\pm}(\Delta; 1-q)\geq D_{\Nn}^{\pm}(\Delta; 1/(1+q))$, so inequality (\ref{eq-main1}) substantially improves such bounds.
  The same is true of the inequality in Theorem 1 below 
which is actually the key to the bounds
(\ref{eq-main1}) and (\ref{eq-main2}). 
This crucial improvement follows from  an almost-sure estimate on the growth of
the  generalized eigenfunctions in the Weyl representation ({\sl cf.}
Proposition \ref{prop-hermiteW} below) which in turn follows from
number-theoretic estimates. As in \cite{KL}, a  control on the
asymptotics of the generalized eigenfunctions then leads to an
improved lower bound on the diffusion coefficients (here by a factor
2 at $q=0$).  }
\end{rem}

\begin{rem}
{\rm 
The bound (\ref{eq-main1}) is of practical interest especially
if $H$ is invariant under some symmetry. Non-symmetric Hamiltonians may lead 
to ballistic motion and absolutely continuous spectral measures (as it is
generically the case for the non-critical Harper Hamiltonian, see \cite{Jit} and
references therein). In this situation, the bound becomes trivial
because both sides in (\ref {eq-main2}) are equal to $1$.
}
\end{rem}

\begin{rem}
{\rm 
Numerical results \cite{TK,RP} as well as the Thouless property \cite{RP}
support that $D_\Nn(-1)=\frac{1}{2}$  in the case of the critical Harper 
Hamiltonian $H_4$ for Diophantine
$\theta/(2\pi)$. According to (\ref{eq-main2}), one thus expects
$\beta_\oned(H_4,\RM;2)\geq\frac{1}{2}$. 
}
\end{rem}

\begin{rem}
{\rm 
Numerical simulations by Pi\'echon \cite{Pie} for the Harper model with  
some strongly incommensurate $\theta/(2\pi)$ indicate that (\ref{eq-main2}) 
may actually  be an exact estimate. Piechon also 
gave a perturbative argument supporting the equality 
$\beta_\oned(H;q)=D_{\Nn}(1-q)$
in the case of the Fibonacci Hamiltonian, and verified it numerically. 
The techniques of the present  article do not apply to the Fibonacci model 
which has no phase-space symmetry. 
}
\end{rem}

\begin{rem}
{\rm 
Our proof forces $q\in(0,1)$ (see Lemma \ref{lem-BGT}). If
$D_\Nn^+(\Delta;q)=D_\Nn^-(\Delta;q)$ for all $q\neq 1$, 
the large deviation technique of \cite{GSB3} leads to 
(\ref{eq-main1}) for all $q>0$ (if $H\in
\Cc^\infty(\Ath)$) and (\ref{eq-main2}) for all $q\in (0,2]$.
Numerical results \cite{TK,RP} suggest that 
the upper and lower fractal dimensions indeed 
coincide for Diophantine $\theta/(2\pi)$. This is hardly to be expected 
for Liouville $\theta/(2\pi)$: the
study in \cite{Las2} can be taken as an indicator for such {\sl
bad} scaling behavior. 
} 
\end{rem}

\begin{rem}
{\rm Two-sided time averages are used for
technical convenience. 
}
\end{rem}

Important intermediate steps of the proof are summarized below.
Associated with the
symmetry $S$ there is a harmonic oscillator Hamiltonian  
$\HGS$ invariant under $\etah_S$ with ground state
$\fis\in\Ss(\RM)$, see
Section~\ref{ssec-symmetries}. In the case of $S_4$ (relevant to the
critical Harper model) this is the conventional harmonic oscillator 
hamiltonian $\HGS_4= (P^2+Q^2)/2$, and $\fis$ is the gaussian state. Let 
$\rhs$ be the spectral measure of $H_W=\piw (H)$ 
with respect to $\fis$. 

\begin{proposi}
\label{prop-phi2dos}
Let $\theta>2\pi$.
There are two positive constants $c_{\pm}$ such that for any Borel subset 
$\Delta \subset \RM$
$$c_- \,\Nn(\Delta) \;\leq \;
  \rhs (\Delta) \;=\;
   \langle \fis | \chi_{\Delta}(H_W)|\fis \rangle\;\leq\;
    c_+\, \Nn (\Delta)\,.
$$

\noindent In particular, $\Nn$ and $\rhs$ have same multifractal exponents.
\end{proposi}

The Hamiltonian $\HGS$ will be used to study transport in phase 
space. Similarly to  eqs.~(\ref{eq-exponent2D}) and (\ref{eq-moment2D}),
moments of the phase space distance
and growth exponents thereof can be defined in the Weyl representation 
as follows: 
$$M_W(H,\Delta;q,t) \;=\;
   \langle \fis |\chi_{\Delta}(H_W) \,
     e^{\imath t H_W }\HGS^{q/2} e^{-\imath t H_W}
      \chi_{\Delta}(H_W) |\fis \rangle \,,
$$
$$\beta_W^{\pm} (H,\Delta;q) \;=\;
   {\lim_{T\to \infty}}^{\!\!\pm} \;
    \frac{\log(\langle M_W(H,\Delta;q,\cdot)\rangle_T)}{q \log(T)}\,.
$$

\begin{proposi}
\label{prop-betamagn}
Let $\theta>2\pi$ and $H=H^*\in \Cc^2(\Ath)$. For $q\in (0,2]$, 
$$\beta_W^{\pm} (H,\Delta;q) \;=\;
   \beta_\twod^{\pm} (H,\Delta;q)\,.
$$
\end{proposi}

\begin{proposi}
\label{prop-beta2gamma}
Let $\theta>2\pi$ and  $H=H^{\ast}\in\Cc^2 (\Ath)$ be invariant under 
$\etah_{S}$  for some symmetry $S\in SL(2,\ZM)$. Then
$$\beta_W^{\pm} (H,\Delta;q) \;\leq\;
   \beta_\oned^{\pm} (H,\Delta;q)\,,\qquad q\in (0,2]\,.
$$
\end{proposi}

Thanks to Propositions~\ref{prop-phi2dos}, \ref{prop-betamagn} and
\ref{prop-beta2gamma} and since $\theta$ may be replaced by $\theta+2\pi$ without
changing the $1D$ and $2D$-representations,
the Main Theorem is a direct consequence of the following:

\begin{theo}
\label{th-gamma2rho}
Let $H=H^{\ast}\in\Cc^2(\Ath)$ and $\theta>2\pi$ 
be a Roth angle. Then, for any 
Borel subset $\Delta \subset \RM $
$$\beta_W^{\pm} (H,\Delta;q) \; \geq \;
   \,D_{\rhs}^{\pm}(\Delta; 1-q)\,,
   \hspace{2cm} 
    \forall\;\; q\in (0,1)\,.
$$
\end{theo}

The proof of Theorem~\ref{th-gamma2rho} will require two 
technical steps that are worth being mentioned here. The first one
requires some notations. Given a symmetry $S$, let $\Pi_S$ be the
projection onto the $H_W$-cyclic subspace $ \Hh_S \subset \Hh$ of
$\fis$. Using the spectral theorem, there is an isomorphism between
$\Hh_S$ and $L^2 (\RM, d\rhs )$. If $(\phi_S^{(n)})_{ n\in\NM}$
denotes the orthonormal basis of eigenstates of $\HGS$ in $\Hh$,
let $\Phi_{n,S}(E)$ be the representative of $ \Pi_S \,\phi_S^{(n)}$
in $L^2 (\RM, d\rhs )$. Then:

\begin{proposi}
\label{prop-hermiteW}
Let $H=H^{\ast}\in\Cc^2(\Ath)$ and let
$\theta$ be a Roth angle. Then for any $\epsilon >0$ there is 
$c_{\epsilon}>0$ such that
$$\sum_{n=0}^{\infty}\,
   |\Phi_{n,S}(E)|^2 \,
    e^{-\delta \, (n+1/2)} \;\leq \;
     c_{\epsilon}\, \delta^{-(1/2 + \epsilon)}\,,
\hspace{1cm}
 \forall \;\; 0 <\delta <1 \,,\qquad \rhs-a.\,e.\;\; E\in\RM\,.     
$$
\end{proposi}

\begin{rem}
{\em This result is uniform ($\rhs$-almost surely) with respect to the spectral
parameter $E$ and to $\delta$. In particular, integrating over $E$ with respect
to $\rhs$ shows that $\sum_{n=0}^{N-1}\,\| \Pi_S
\, \phi_S^{(n)} \|^2 = O(N^{1/2+\epsilon})$. This is possible because
of the following complementary result proved in the Appendix:
}
\end{rem}

\begin{proposi}
\label{prop-multiplicity}
Let $H=H^{\ast}\in\Ath$. Then $H_W$ has infinite multiplicity and
no cyclic vector.
\end{proposi}

The second technical result 
concerns the so-called {\em Mehler kernel} of the Hamiltonian $\HGS$,  
notably the integral kernel of the operator $e^{-t\, \HGS}$ in the
$Q$-representation:
\begin{equation}
\label{eq-mehler}
\Mm_S (t;x,y)\;=\;
   \langle x |\, e^{-t\, \HGS}\,|y\rangle \,,
\end{equation}

\begin{proposi}
\label{prop-mehler}
Let $\theta$ be a Roth angle. Then, for all $\epsilon>0$,
$$\sup_{0\leq x\leq 2\pi\theta^{-1/2},\, 0\leq y\leq \theta^{1/2}}\;
   \sum_{m\in\ZM^2}\,
    |\Mm_S (t;x+2\pi m_1\theta^{-1/2},y + \theta^{1/2} m_2)|\; =\;
     \Oo(t^{-1/2-\epsilon})\,,
\hspace{1cm}
  as\;\; t \downarrow 0\,.
$$
\end{proposi}


\vspace{.2cm}

\section{Weyl's calculus}
\label{sec-weylcalculus}

This chapter begins with a review of basic facts  about 
Weyl operators, the rotation algebra and implementation of symmetries
therein. The formulas are well-known ({\sl e.g.} \cite{Per,Bel94}
and mainly given in order to fix notations, but for the convenience of the
reader their proofs are nevertheless given in the Appendix. The
chapter also contains a new and compact 
solution of the frame problem for coherent states (Section
\ref{ssec-thetatraceframe}).

\subsection{Weyl operators}
\label{ssec-weyloperators}

\noindent Let $\Hh$ denote the Hilbert space $L^2(\RM)$. 
Given a vector $\aG=(a_1,a_2)\in
\RM^2$, the associated Weyl operator is defined  by:
\begin{equation}
\label{eq-Weyldef}
\WG (\aG)\,=\,
   e^{\imath (a_1 P + a_2 Q)}
   \hspace{.7cm} \Leftrightarrow \hspace{.7cm}
   \WG (\aG) \,\psi (x) \,=\, 
    e^{\imath a_1 a_2/2}\, e^{\imath a_2 x}\,
     \psi (x+ a_1)\,,
      \hspace{.7cm}
       \forall\;\;\psi \in \Hh\,.
\end{equation}

\noindent The Weyl operators are unitaries, strongly continuous with
respect to $\mathfrak{a}$ and satisfy
\begin{equation}
\label{eq-Weylccr}
\WG (\aG)\,\WG (\bG)\,=\,
   e^{\imath \,\aG\wedge \bG/2}\;\WG (\aG +\bG)\,,
    \hspace{1cm}
     \aG\wedge \bG \,=\, a_1 b_2 - a_2 b_1\,.
\end{equation}

\noindent The following weak-integral 
identities are verified in the Appendix:

\begin{equation}
\label{eq-WeylHusimi}
\langle \psi |\WG (\aG)^{-1} \,|\psi \rangle\; \WG (\aG) \ =\;
 \int_{\RM^2}\, \frac{d^2 \bG}{2\pi}\,
  e^{\imath \,\aG\wedge \bG}\;
   \WG (\bG)\,|\psi\rangle \langle \psi | \WG (\bG)^{-1}\;,
\end{equation}
\begin{equation}
\label{eq-HusimiWeyl}
\WG (\bG)\,|\psi\rangle \langle \psi | \WG (\bG)^{-1}\;=\;
 \int_{\RM^2}\, \frac{d^2 \aG}{2\pi}\,
  e^{\imath \,\bG\wedge \aG}\;
  \langle \psi |\WG (\aG)^{-1} \,|\psi \rangle\; \WG (\aG) \;.
\end{equation}

\noindent Applying (\ref{eq-WeylHusimi}) to $\phi$ and setting $\aG=0$
leads to

\begin{equation}
\label{eq-cyclicity}
\phi \, =\, 
     \int_{\RM^2}\, \frac{d^2 \bG}{2\pi}\,\;
      \langle \psi | \WG(\bG)^{-1}| \phi \rangle\;
       \WG (\bG) \, \psi\,
\mbox{ , }
\qquad
\phi,\; \psi\, \in\,\Hh\, \mbox{ , }\qquad \| \psi\| =1\,
\mbox{ . }
\end{equation}

\noindent In particular, any non zero vector in $\Hh$ is cyclic for the
Weyl algebra $\{\WG(\aG)|\aG\in\RM^2\}$. 
If $\psi \in \Hh$, the map $\aG\in \RM^2 \mapsto
\langle \psi |\WG (\aG) |\psi \rangle\in\CM$ is continuous, tends to
zero at infinity and belongs to $L^2 (\RM^2)$, whereas $\psi \in \Ss(\RM)$
if and only if this map belongs to $\Ss(\RM^2)$.

\subsection{The rotation algebra}
\label{ssec-rotationalg}

The {\em rotation algebra} $\Ath$, its representations
$(\pi_{\omega})_{\omega \in \RM}$, $\pi_\twod$ and $\piw$ as well as
the tracial state $\TV_{\theta}$ and $\ast$-derivations  $\delta_1,\delta_2$
were defined in Section~\ref{sec-notresults}. Here we give some
complements, further definitions and the short proof of Proposition
\ref{prop-multiplicity}. 
The  trace is faithful and satisfies the Fourier formula:
\begin{equation}
\label{eq-AthFourier}
A\,=\, 
  \sum_{\lL\in\ZM^2}\, a_{\lL}\, \Wth (\lL)\,,
\qquad
    a_{\lL} \,=\, \TV_{\theta} (\Wth (\lL)^{-1} \,A)\,
\mbox{ . }
\end{equation}

\noindent In addition, 
\begin{equation}
\label{eq-trace1D}
\TV_{\theta} (A) \;=\;
   \int_{}^{2\pi} \, \frac{d\omega}{2\pi} \,
    \langle m | \pi_{\omega} (A) |m\rangle \,
    \;=\;
    \langle \lL | \pi_\twod (A) |\lL\rangle \,,
     \hspace{1cm} \forall\; A \in \Ath\,,\;\forall\; 
      m\in\ZM,\;\forall\; \lL\in\ZM^2\,.
\end{equation}

\noindent The $\ast$-derivations satisfy 
$\delta_j\, \Wth (\mM)=\imath\, m_j \Wth (\mM)$, $j=1,2$.
It follows from (\ref{eq-AthFourier})
that $A\in\Cth$ if and only if the sequence of its Fourier coefficients
is fast decreasing. If $A\in\Cth$ and $A$ is invertible in $\Ath$, then
$A^{-1}\in\Cth$. The position operator $(X_1,X_2)$ defined on the space
$\Sssmall(\ZM^2)$ of Schwartz sequences in $\ell^2(\ZM^2)$ forms a 
{\em connection} \cite{Co} in the following sense  
\begin{equation}
\label{eq-connectiontwod}
X_j (\pi_\twod(A)\phi)\;=\;
   \pi_\twod (\delta_j A)\phi +\pi_\twod(A)X_j\psi\,
    \hspace{1cm}
     \forall \;\;A\in\Cth\,,\qquad \phi\in\Sssmall(\ZM^2)\,.
\end{equation}

\noindent Similarly, if $(\nabla_1, \nabla_2)$ is defined on $\Ss(\RM)$
by $\nabla_1=-\imath Q/\sqrt{\theta},\; \nabla_2=\imath P/\sqrt{\theta}$, then 
 
\begin{equation}
\label{eq-connection}
\nabla_j (\piw(A)\psi)\;=\;
   \piw (\delta_j A)\psi +\piw(A)\nabla_j\psi\,
    \hspace{1cm}
     \forall \;\;A\in\Cth\,,\qquad \psi\in\Ss(\RM)\,.
\end{equation}

\noindent Then $\Ss(\RM)$ is exactly the set of $\Cc^{\infty}$-elements of $\Hh$
with respect to $\naV$. In particular, if $\psi\in\Ss(\RM)$ and $A\in\Cth$, then
$\piw(A)\psi\in\Ss(\RM)$.

\vspace{.2cm}

For the Weyl representation, let us use the notations

\begin{equation}
\label{eq-Weylredef}
\piw (\Wth (\mM))
\;=\;\Wwth(\mM)
\;:=\; \WG(\sqrt{\theta}\mM)
\mbox{ , }
\qquad
\forall\;\;\mM\in\ZM^2
\mbox{ . }
 \end{equation}

\noindent It can be seen as a direct integral of 
$1D$-representations by introducing the family $(\Gg_{\omega})_{\omega\in
\RM}$ of transformations from $\Hh$ into $\ell^2(\ZM)$
\begin{equation}
\label{eq-geeomega}
(\Gg_{\omega} \phi)(n)\,=\,
   \theta ^{-1/4} \,
    \phi \, \left(
             \frac{\omega - n\theta}{\sqrt{\theta}}
            \right)\,,
     \hspace{.7cm} \forall\;\; \phi\in\Hh\,
\mbox{ . }
\end{equation}
\noindent Then a direct computation (given in the Appendix) shows that:
\begin{equation}
\label{eq-intdirrep}
\langle \phi |\piw (A)| \phi \rangle\; =\;
    \int_{0}^{\theta} \, d\omega\,
     \langle \Gg_{\omega} \phi| \pi_{\omega} (A) |\Gg_{\omega} \psi
     \rangle\,
\mbox{ , }
A\in \Ath\mbox{ , }
\phi,\psi\in\Hh
\mbox{ . }
\end{equation}
\noindent In particular, $\| \phi \|^2 \, =\,
\int_{0}^{\theta} \, d\omega\,\| \Gg_{\omega} \phi\|^2_{\ell^2}$.
The link between $\piw$ and $\pi_\twod$ will be established
in Section \ref{ssec-magnetic}.

\vspace{.2cm}

It follows from a theorem by Rieffel \cite{Rie} that the 
commutant of $\piw (\Ath)$ is the von Neumann algebra generated by 
$\piw (\Athp)$ where $\theta'/2\pi = 2\pi/\theta$ 
and $\piw (\Wthp (\lL))\,=\, \Wwthp (\lL) $. 
The following result is proven  in the Appendix:

\begin{proposi}
\label{prop-Poissonsum} {\rm (The
generalized Poisson summation formula):}

\begin{equation}
\label{eq-PoissonforT}
\Tpsi\;:=\;
 \sum_{\lL\in\ZM^2}\,
  \Wwthp (\lL) \,|\psi\rangle \, \langle \psi| \Wwthp (\lL)^{-1}\;=\;
   \frac{\theta}{2\pi}
    \sum_{\mM\in\ZM^2}\,
     \langle \psi | \Wwth (\mM)^{-1}|\psi\rangle\, \Wwth (\mM)\;.
\end{equation}

\end{proposi}

By eq.~(\ref{eq-PoissonforT}),
$\psi\in\Ss(\RM)$ implies $\Tpsi \in \Cth$.
It follows immediately from eq.~(\ref{eq-PoissonforT}) that, given
$\psi\in\Ss(\RM)$, there is a positive element in $\Ath$, denoted $\Dpsi$, 
such that $\Tpsi
=(\theta/2\pi)\;\piw\left(\Dpsi\right)$.
Moreover
\begin{equation}
\label{eq-psipsi}
\langle \psi |\,\piw (A) \,|\psi\rangle \;=\;
 \TV_{\theta} \left(
   A\, \Dpsi
              \right)\,,
    \hspace{1cm}
     \forall A\in\Ath\,.
\end{equation}

\vspace{.2cm}

\subsection{Symmetries}
\label{ssec-symmetries}

It is well-known 
that $S\in SL(2,\RM)$ can be uniquely decomposed in a torsion,
a dilation and a rotation as follows :

$$S \;=\; 
   \left(
    \begin{array}{cc}
    a & b\\
    c & d
    \end{array}
   \right) \;=\;
    \left(
     \begin{array}{cc}
     1 & 0\\
     \kappa & 1
     \end{array}
    \right) \;
      \left(
       \begin{array}{cc}
        \lambda & 0\\
        0 & \lambda^{-1}
        \end{array}
       \right) \;
        \left(
         \begin{array}{cc}
          \cos{s} & -\sin{s}\\
           \sin{s} & \cos{s}
          \end{array}
        \right) \;,
$$

\noindent with $\kappa=(ac+db)/(a^2+b^2),\; \lambda=(a^2+b^2)^{1/2},\; 
e^{\imath s}= (a-\imath b)(a^2+b^2)^{-1/2}$. Moreover, 
if $S\in SL(2,\RM)$, then 
there is a unitary transformation $\Ff_S$ acting on $\Hh$
such that
\begin{equation}
\label{eq-weylimp}
\WG (S\aG) \;=\;
   \Ff_S\, \WG(\aG)\, \Ff_S^{-1} \,,
\qquad \aG\in\RM^2\;,
\end{equation}

\noindent as shows the above decomposition as well as
the following result, the proof of which is deferred to the Appendix:

\begin{proposi}
\label{prop-sl2rweyl}
For any $\kappa,\lambda,s\in\RM$, $\lambda\neq 0$, up to a phase

\begin{equation}
\label{eq-sl2rweyl}
\Ff_{\mbox{\tiny 
     $
      \left(\begin{array}{cc}
     1 & 0\\
     \kappa & 1
     \end{array}\
     \right)$
     }}=
      e^{-\imath\,\kappa\,Q^2/2}\,,
\hspace{.3cm}
\Ff_{\mbox{\tiny 
     $
      \left(\begin{array}{cc}
        \lambda & 0\\
        0 & \lambda^{-1}
     \end{array}\right)$
     }}=
      e^{-\imath\,\ln({\lambda})\,(QP+PQ)/2}\,,
\hspace{.3cm}
\Ff_{\mbox{\tiny 
     $
      \left(\begin{array}{cc}
          \cos{s} & -\sin{s}\\
           \sin{s} & \cos{s}
     \end{array}\right)$
     }}=
      e^{-\imath\,s \,(Q^2+P^2-1)/2}\,.
\end{equation}

\noindent Note in particular that  
$\Ff_S\, \Ff_{S'} = z\,\Ff_{S\,S'}$ for $z\in\CM\,,|z|=1$. 
Furthermore, if $0< s<\pi$,
\begin{equation}
\label{eq-frotation} 
\Ff_{\mbox{\tiny 
     $\left(\begin{array}{cc}
          \cos{s} & -\sin{s}\\
           \sin{s} & \cos{s}
     \end{array}\right)$
     }}\,\phi (x)\;=\;
    \int_{\RM}\; \frac{dy}{\sqrt{2\pi\,\sin{s}}}\;
     e^{\imath 
      \left(
       \cos{s}\, (x^2+y^2)-2xy
      \right)/2\sin{s}
     }\; \phi (y)\;.
\end{equation}
\end{proposi}

\vspace{.2cm}

In the special case $s=\pi/2$, namely for the matrix $S_4$ (see
Section~\ref{sec-notresults}), this gives the usual Fourier transform 
\begin{equation}
\label{eq-fourier4}
\Ff_{S_4}\,\phi (x)\;=\;
   \int_{\RM}\, 
    \frac{dy}{\sqrt{2\pi}}\,
     e^{-\imath xy}\,
      \phi (y)\,.
\end{equation}

\noindent For the case of the $3$-fold and $6$-fold symmetries $S_3$ and $S_6$,
acting on a hexagonal or a triangular lattice (see
Section~\ref{sec-notresults}), eqs. 
(\ref{eq-sl2rweyl}) and (\ref{eq-frotation}) give
\begin{equation}
\label{eq-fourier3-6}
\Ff_{S_3}\,\phi (x)\;=\;
  e^{\imath \pi/12}\,
   \int_{\RM}
    \frac{dy}{\sqrt{2\pi}}\,
     e^{-\imath x(x+2y)/2}\,
      \phi (y)\,,
\hspace{.7cm}
  \Ff_{S_6}\,\phi (x)\;=\;
  e^{-\imath \pi/12}\,
   \int_{\RM}
    \frac{dy}{\sqrt{2\pi}}\,
     e^{-\imath y(2x-y)/2}\,
      \phi (y)\,.
\end{equation}

Now suppose that $S\in SL(2,\RM)$ satisfies $S^r={\bf 1}$ for some
$r\in\NM$, $r\geq 2$ and $S^n\neq {\bf 1}$ for $n<r$.
It will be convenient to introduce the following operator acting on $\Hh$
$$\HGS \;=\;
   \frac{1}{2r}\,\sum_{n=0}^{r-1}\,
    \Ff_{S}^n\, Q^2 \Ff_{S}^{-n}\;=\;
     \frac{1}{2}\;\langle K|M_S| K\rangle\,,
\hspace{1cm}
     M_S\;=\;
      \frac{1}{r}\,\sum_{n=0}^{r-1}
       S^n\, |e_2\rangle \langle e_2| \, (S^t)^n\,,
$$

\noindent where $K= (P,Q)$ and $\{ e_1,e_2\}$ is the canonical basis of
$\RM^2$. Note that ${\mathfrak H}_{S_4}=(P^2+Q^2)/2$.
There is $0\leq n\leq r-1$ such that 
$S^n e_2\wedge e_2 \neq 0$, 
so $M_S$ is positive definite and can be diagonalized by a rotation:

$$
M_S
\;=\;
\left( \begin{array}{cc}\cos{\gamma} & -\sin{\gamma}\\
\sin{\gamma} & \cos{\gamma}\end{array} \right) 
\left( \begin{array}{cc}\mu_S^+ & 0 \\
0 & \mu_S^-\end{array} \right) 
\left( \begin{array}{cc}\cos{\gamma} & -\sin{\gamma}\\
\sin{\gamma} & \cos{\gamma}\end{array} \right)^{-1} 
\mbox{ . }
$$

\noindent 
Hence $\HGS$ is unitarily equivalent to the harmonic oscillator Hamiltonian 
$(\mu_S^{+}P^2 +\mu_S^{-}Q^2)/2$. Therefore, 

\begin{equation}
\label{eq-oscillatorS}
\HGS\;=\;
      \mu\,\sum_{n=0}^{\infty} \,
       \left(n+\frac{1}{2}\right)
        \,|\phi_S^{(n)}\rangle \langle \phi_S^{(n)}|\,,
\qquad
\mu\;=\;(\mu_S^+\mu_S^-)^{1/2} \mbox{ , }
\qquad
\lambda\;=\;\left(\frac{\mu_S^+}{\mu_S^-}\right)^{1/4}
\mbox{ , }
\end{equation}

\noindent where the $\phi_S^{(n)}$ are the eigenstates. The
ground state is denoted $\fis\equiv\phi_S^{(0)}$. 

\begin{proposi}
\label{prop-mehlerS}
Up to a phase, the ground state is given by
\begin{equation}
\label{eq-groundstate}
\fis (x) \;=\; 
   \left(
    \frac{\Re e{(\sigma_S)}}{\pi}
   \right)^{1/4}\;
    e^{-\sigma_S\,x^2/2}\,,
\hspace{1cm}
  \sigma_s\;=\;
   \frac{
    \sqrt{\mu_S^-}\,\cos{\gamma} +\imath \sqrt{\mu_S^+}\sin{\gamma}
        }{
    \sqrt{\mu_S^+}\,\cos{\gamma}+ \imath \sqrt{\mu_S^-}\sin{\gamma}
          }
\mbox{ , }
\end{equation}

\noindent and the Mehler kernel 
{\rm (\ref{eq-mehler})} by

\begin{equation}
\label{eq-mehlerS}
\Mm_S (t;x,y)\;=\;  
 \frac{e^{-
   \frac{(x-y)^2 \tanh{(t\mu)}^{-1}+ (x+y)^2 \tanh{(t\mu)}}
   {4(\lambda^2 \cos{\gamma}^2 + \lambda^{-2} \sin{\gamma}^2)} }}
  {\lambda \sqrt{2\pi\sinh{(t\mu)}
  (\lambda^2 \cos{\gamma}^2 + \lambda^{-2} \sin{\gamma}^2)}}\;\;
    e^{\imath (x^2-y^2)
     \frac{\sin{(2\gamma)}(\lambda^2-\lambda^{-2})}
      {4(\lambda^2 \cos{\gamma}^2 + \lambda^{-2} \sin{\gamma}^2)}}\,.
\end{equation}

\end{proposi}

\vspace{.2cm}

By construction,
$\Ff_S \HGS \Ff_S^* = \HGS$, so that $\Ff_S \fis= e^{\imath \delta_S} \fis$
for some phase $\delta_S$. Thus, it is possible to choose the phase of $\Ff_S$
such that $\Ff_S \fis= \fis$. Such is the case for $\Ff_{S_i}$ in
eqs.~(\ref{eq-fourier4}) and  (\ref{eq-fourier3-6}). 

\vspace{.2cm}

Recall from Section~\ref{sec-notresults} that
$\pm \1\neq S\in SL(2,\ZM)$ is called a symmetry of $\Ath$ if
$\sup_{n\in\ZM}\|S^n\| <\infty$.
Since the set of $M\in SL(2,\ZM)$ with $\|M\|\leq c$ is finite (for any
$0<c<\infty$), and since $S\neq \pm \1$, there 
is an integer $r\in\NM_{\ast}$ such
that $S^r =\1$ and $S^n\neq \1$ for $0<n<r$. So the two eigenvalues are
$\{ e^{\pm\imath\varphi_s} \}$, with $r\varphi_s =0 \pmod{2\pi}$ and 
$\varphi_s
\neq 0,\pi$. In particular $\tr (S) = 2\cos{\varphi_s}\in\ZM$, implying $r\in
\{ 3,4,6\}$ and $\varphi_s \in \{ \pm \pi/3\,, \pm \pi/2 \,,\pm 2\pi/3 \}$.
Any $S\in SL(2,\ZM)$ defines a $\ast$-automorphism $\etah_S$ of $\Ath$
through $\etah_S (\Wth (\mM))=\Wth (S \mM)$. According to the above,
$\piw(\etah_S (\Wth (\mM)))=\Ff_S\piw(\Wth (S \mM))\Ff_S^{-1}$.

\vspace{.2cm}

\subsection{$\theta$-traces and $\theta$-frames}
\label{ssec-thetatraceframe}

\begin{defini}
\label{def-trace}
A vector $\psi \in \Hh$ will be called {\em $\theta$-tracial} if
$\langle \psi |\Wwth (\lL)\, |\psi \rangle = \TV_{\theta}(\Wth (\lL))
= \delta_{{\bf l},0}$ for all
$\lL\in\ZM^2$. Equivalently, the family $(\Wwth(\lL)\psi)_{\lL\in\ZM^2}$ is
orthonormal. 
\end{defini}

Using the commutation rules (\ref{eq-Weylccr}), it is possible
to check that $\psi$ is $\theta$-tracial if and only if $\WG (\aG) \psi$
is $\theta$-tracial for any $\aG \in\RM^2$. It also follows from
eq.~(\ref{eq-PoissonforT}) that $\psi$ is $\theta$-tracial if and only if
$\Tpsi = (\theta/2\pi)\1$. Such  $\theta$-tracial states
exist under the following condition: 

\begin{theo}
\label{theo-tracialvectors}
There is a $\theta$-tracial vector $\psi\in\Hh$ if and only if $\theta\geq
2\pi$. If $\theta>2\pi$ there is a $\theta$-tracial vector in $\Ss (\RM)$.
For $\theta \geq 2\pi$, denote by $\Pi_\psi$ the projection
on the orthocomplement of the $\psi$-cyclic subspace 
$\overline{\piw(\Ath)\psi}\subset \Hh$. There is a projection
$P_\psi\in\Athp$ satisfying $\piw(P_\psi)=\Pi_\psi$ and 
$\TV_{\theta'}(P_\psi)= 1-2\pi/\theta$.
In particular, $\psi$ is also $\Ath$-cyclic for $\theta =2\pi$.
\end{theo} 

\noindent {\bf Proof: } If $\psi$ is $\theta$-tracial, then $(\theta/2\pi) =
\langle \psi|\Tpsi|\psi\rangle  =
\sum_{\lL\in\ZM^2} \, |\langle \Wwthp (\lL)\psi \,|\psi\rangle |^2
\geq \|\psi\|^2 =1$.

If $\theta >2\pi$, for $0<\varepsilon <\min{(2\pi,\theta- 2\pi)}$,
let $\phi$ be a $C^{\infty}$ function on $\RM$ such that $0\leq \phi
\leq 1$,
with support in $[0,2\pi+\varepsilon]$, such that $\phi =1$ on
$[\varepsilon, 2\pi ]$, and $\phi(x)^2 + \phi(x+2\pi)^2 =1$ whenever
$0\leq x \leq \varepsilon$. Using (\ref{eq-intdirrep}), $\phi$
is $\theta$-tracial (after normalization), and belongs to $\Ss(\RM)$.
If $\theta =2\pi$, the same argument holds with $\varepsilon =0$.
Then $\phi\in\Hh$, but it is not smooth anymore.

Let $\psi$ be $\theta$-tracial. Exchanging the r\^oles of $\theta$
and $\theta'$, the Poisson summation formula implies
$${\Tpsi}'\,=\,
   \sum_{\mM\in\ZM^2}\,
    \Wwth (\mM) |\psi\rangle \langle \psi|\Wwth (\mM)^{-1}\;=\;
     \frac{2\pi}{\theta}\,
      \sum_{\lL\in\ZM^2}\,
       \langle \psi|\Wwthp (\lL)^{-1}|\psi\rangle\,
        \Wwthp(\lL)\,.
$$

\noindent Hence $\Pi_\psi=1-{\Tpsi}'$ is the desired orthonormal
projection which, due to the r.h.s., is the Weyl representative of an
element $P_\psi\in \Athp$. Its trace
is $\TV_{\theta'}(P_\psi)= 1-2\pi/\theta$. If $\theta
=2\pi$, since the trace is faithful, ${\Tpsi}'=\1$, so that $\psi$ is
cyclic.
\hfill $\Box$

\vspace{.2cm}

\begin{defini}
\label{def-frame}
A vector $\psi\in\Hh$ is called a {\em $\theta'$-frame}, if there
are constants $0<c<C <\infty$ such that $c\1\leq \Tpsi \leq C\,\1$. 
\end{defini}

This definition is in accordance with the literature (\cite{Sei} and
references therein) where the complete set $(\Wwthp(\lL)\psi)_{\lL\in\ZM^2}$
is called a frame. The principal interest of frames is due to the
following: any vector $\phi\in\Hh$ can be decomposed as
$\phi=\Tpsi {(\Tpsi)}^{-1} \phi=\sum_{\lL\in\ZM^2} c_\lL\,\Wwthp(\lL)\psi$
where $c_\lL=\langle \psi|\Wwthp(\lL)^*{(\Tpsi)}^{-1}|\phi\rangle$. 
If $\psi\in\Ss(\RM)$ and $\phi\in\Ss(\RM)$, then $(c_\lL)_{\lL\in\ZM^2} 
\in \Sssmall(\ZM^2)$. 
Further note that, if $\psi$ is a $\theta'$-frame, then 
$\hat{\psi} = (\theta/2\pi)^{1/2}{(\Tpsi)}^{-1/2}\psi$ 
is $\theta$-tracial. In addition, if
$\psi\in \Ss(\RM)$ then $\hat{\psi} \in \Ss(\RM)$. 

\vspace{.2cm}

The next result shows that so-called
Weyl-Heisenberg or Gabor lattices constructed with a gaussian mother
state are frames if only the volume of the chosen phase-space cell is
sufficiently small. This was proved in \cite{Sei}, but the present 
proof is new and covers more general cases.

\vspace{.2cm}
 
Suppose $S\in SL(2,\RM)$ satisfies $S^r=1$ for some $r$. 
Using the results of Section~\ref{ssec-symmetries} and
eq.~(\ref{eq-Weyldef}), it is possible to compute
\begin{equation}
\label{eq-gausstate}
\langle \fis |\WG (\aG)| \fis \rangle \;=\;
 e^{-|\aG\,|_S^2/4}\;,
\hspace{1cm}
  |\aG\,|_S^2 \;=\;
   \frac{\mu_S^+\,a_1^2 + \mu_S^-\,a_2^2}{\mu}\;.
\end{equation}

\begin{theo}
\label{theo-grounstate}
For $\theta > 2\pi$, $\fis$ is a $\theta'$-frame in $\Ss(\RM)$. 
\end{theo} 

\noindent {\bf Proof: } The proof below is given for $\fio\equiv\phi_{S_4}
$, but the same strategy works for any $\fis$.

Thanks to Poisson's formula (\ref{eq-PoissonforT})
and eq.~(\ref{eq-gausstate}), $\Tphio\leq (\theta/2\pi)\sum_{\mM}
e^{-\theta|\mM|^2/4}$. It is therefore enough to find a positive lower bound.
Since $\piw$ is faithful, it is enough to show that $T_0 = \sum_{\mM}
e^{-\theta|\mM|^2/4} \Wth(\mM)$ is itself bounded from below in $\Ath$. Writing
$\theta = 2\pi +\delta$ with $\delta >0$, there is a $\ast$-isomorphism between
$\Ath$ and the closed subalgebra of $\Aa_{2\pi}\otimes \Aa_{\delta}$ generated
by $(W_{2\pi}(\mM)\otimes W_{\delta}(\mM))_{\mM\in\ZM^2}$. It is enough
to show that $\hat{T}_0 = \sum_{\mM} e^{-\theta|\mM|^2/4} W_{2\pi}(\mM)\otimes
W_{\delta}(\mM)$ is bounded from below in $\Aa_{2\pi}\otimes \Aa_{\delta}$.
$\Aa_{2\pi}$ is abelian and
$\ast$-isomorphic to $C(\TM^2)$, provided $W_{2\pi}(\mM)$ is identified with
the map $\kaK = (\kappa_1,\kappa_2)\in\TM^2\mapsto (-1)^{m_1 m_2} e^{ \imath
\kaK \cdot \mM}\in\CM$. Hence it is enough to show that $\hat{T}_0 (\kaK) =
\sum_{\mM} (-1)^{m_1 m_2}\, e^{-\theta|\mM|^2/4\,+\imath \kaK \cdot \mM}\,
W_{\delta}(\mM)$ is bounded from below in $\Aa_{\delta}$ uniformly in $\kaK$.
Since the Weyl representation is faithful, $W_{\delta}(\mM)$ can be replaced 
by $\Ww_{\delta}(\mM)$. Using eq.~(\ref{eq-WeylHusimi}) with $\psi=\fio$ and
$\aG= \sqrt{\delta} \mM$, it is thus enough to show that
$$
\tilde{T}_0 (\kaK)\;=\;
   \int_{\RM^2}\;
    \frac{d^2 \bG}{2\pi}\,
      \Theta(\kappa_1+\sqrt{\delta} b_2, \kappa_2-\sqrt{\delta} b_1)\,
       \WG(\bG)|\fio\rangle \langle \fio | \WG(\bG)^{-1} \,,
$$
where
\begin{equation}
\label{eq-Fdef}
\Theta(\kaK)\;=\;
\sum_{\mM\in\ZM^2} (-1)^{m_1 m_2}\, e^{-\pi|\mM|^2/2\,+\imath (\kaK \cdot
\mM\,)}\;,
\end{equation}

\noindent is bounded from below. Clearly the function $\Theta$ is
$2\pi$-periodic in both of its arguments. Hence, decomposing the
integral into a sum of integrals over the shifted unit cell
$C=[0,2\pi)\times[0,2\pi)$ and using $\Ww_{\delta'}(\aG)=\WG(2\pi
\aG/\sqrt{\delta})$ gives
$$
\tilde{T}_0 (\kaK)\;=\;
  \sum_{\lL\in\ZM^2} 
   \int_{C}\;
    \frac{d^2 \aG}{2\pi\delta}\,\Theta(\aG)\,
     \Ww_{\delta'}\left(\lL+\frac{\aG+\hat{\kappa}}{2\pi}\right)
      |\fio\rangle \langle \fio | 
       \Ww_{\delta'}\left(\lL+\frac{\aG+\hat{\kappa}}{2\pi}\right)^{-1} \,,
$$

\noindent where $\hat{\kappa}=(\kappa_2,-\kappa_1)$. The Poisson
summation formula applied to the summation over $m_1$ in
(\ref{eq-Fdef}) gives a sum over an index $n_1$. Changing summation
indexes $n_2=m_2-n_1$ shows that
$\Theta(\kaK)= \sqrt{2} \,e^{-\kappa_1^2/2\pi}\, |f(\kappa_1+\imath\kappa_2)|^2 $,
where $f$ is the holomorphic entire function given by $f(z)=\sum_{n\in\ZM}\,
e^{-\pi\,n^2\,-nz}$. It can be checked that $f(z+2\imath\pi) = f(z)$ and
that $f(z+2\pi) = e^{z+\pi}\,f(z)$. Moreover, using the Poisson summation
formula, $f$ does not vanish on $\gamma$, the boundary of $C$ 
oriented clockwise. As $\Theta$ has no poles, 
the number of zeros of $f$ within $C$ counted with their multiplicity
is given by $\oint_{\gamma} df/2\imath\pi f$. Using the  
periodicity properties of $f$, this integral equals $1$.
Moreover, a direct calculation shows that the unique zero with
multiplicity 1 of $f$ 
lies at the center $\pi(1+\imath)$ of $C$. 
Hence there is a constant $c_1>0$ such that
$|f(\pi+\imath\pi+re^{\imath\varphi})|\geq c_1 r^2$ for all $\varphi\in
[0,2\pi)$. Let $B_r$ denote the ball of size $r$ around 
$\pi(1+\imath)$. Replacing this shows 

$$
\tilde{T}_0 (\kaK)\;\geq\;
\frac{c_1 r^2}{\delta}
\left({\bf 1}-\int_{B_r} 
\frac{d^2 \aG}{2\pi}\,
     \Ww_{\delta'}\left(\frac{\aG+\hat{\kappa}}{2\pi}\right)
      \;T^{\delta'}_{\fio}\;
       \Ww_{\delta'}\left(\frac{\aG+\hat{\kappa}}{2\pi}\right)^{-1}\right)\;.
$$

\noindent As $T^{\delta'}_{\fio}\leq c_2{\bf 1}$,
$\tilde{T}_0 (\kaK)\geq {\bf 1}\,c_1 r^2(1-c_2 r^2/2)/\delta$.
Choosing $r$ small enough, $\tilde{T}_0 (\kaK)$ is bounded from below 
by a positive constant uniformly in $\kaK$. 
\hfill $\Box$


\vspace{.2cm}

\section{Comparison theorems}
\label{sec-1D2Weyl}

\subsection{Proof of Proposition~\ref{prop-phi2dos} }
\label{ssec-DOS}

Let $H=H^{\ast}\in \Ath$ and set $H_W = \piw (H)$.
For normalized $\phi\in\Hh$, 
$\rho_{\phi}$ denotes the spectral measure of $H_W$ relative to $\phi$.
Proposition~\ref{prop-phi2dos} is a corollary of the following 
result:

\begin{theo}
\label{theo-rhophivsNn}
For $\theta \geq 2\pi$, for any normalized 
$\theta'$-frame $\phi\in\Hh$ and any Borel 
subset $\Delta$ of $\RM$,
\begin{equation}
\label{eq-rhophivsNn}
\frac{2\pi}{\theta}
  \|{(\Tphi)}^{-1}\|^{-1}\,
   \Nn (\Delta)\; \leq \;
    \rho_{\phi}(\Delta) \;\leq \;
     \frac{2\pi}{\theta}
      \|\Tphi\|\,\Nn (\Delta)\,.
\end{equation}

\end{theo} 

\noindent {\bf Proof: } 
Eq.~(\ref{eq-psipsi}) leads to 
$$\rho_{\phi}(\Delta) \;= \;
   \TV_{\theta}
    \left(
     \chi_{\Delta}(H)\,\Dphi 
    \right)\;\leq\;
     \|\Dphi\|\, \Nn(\Delta)\,,
$$

\noindent and to
$$\Nn(\Delta) \;=\;
    \TV_{\theta}
     \left(
      \chi_{\Delta}(H)\,\Dphi{(\Dphi)}^{-1}
     \right)\; \leq \;
      \rho_{\phi}(\Delta)\, \|{(\Dphi)}^{-1}\|\,.
$$
\noindent Since $\Tphi =\theta/2\pi\; \piw(\Dphi)$, the
theorem follows.  
\hfill $\Box$

\subsection{Proof of Proposition~\ref{prop-betamagn} }
\label{ssec-magnetic}

Let $\theta>2\pi$. The ground state $\phi_S$ of $\HGS$ is a
$\theta'$-frame according to Theorem \ref{theo-grounstate}.
Let $\psi_S = (\theta/2\pi)^{1/2}{(T^\theta_{\phi_S})}^{-1/2}\phi_S$ 
be the associated $\theta$-tracial vector. Further set
$\Hh_S=\overline{\piw(\Ath) \psi_S}$. In this section, $\piw$ denotes
the restriction of the Weyl representation to $\Hh_S$. A unitary
transformation $\Uu:\Hh_S\to \ell^2(\ZZ^2)$ is defined by

$$
(\Uu\phi) (\lL)\;=\;\langle\psi_S|
\Wwth(\lL)^{-1}|\phi\rangle
\mbox{ , }
\qquad
\phi\in\Hh_S\mbox{ , }
\;\;\lL\in\ZM^2\mbox{ . }
$$

\noindent Then $\Uu\piw(A)\Uu^*=\pi_\twod(A)$ for all $A\in\Ath$. 
Moreover $\Uu:\Ss(\RM)\cap\HGS\to\Sssmall(\ZM^2)$.
As $\,\Uu \psi_S =|{\bf 0}\rangle$, 

$$
M_W(H,\Delta;q,t)
\;=\;
   \langle {\bf 0} |\chi_{\Delta}(H_\twod) \,
     e^{\imath H_\twod t}(\Uu\HGS \Uu^*)^{q/2} e^{-\imath H_\twod t}
      \chi_{\Delta}(H_\twod) |{\bf 0} \rangle \,.
$$

\noindent Recall that $\HGS$ is a polynomial of second degree in $Q$
and $P$. From (\ref{eq-connection}) follows

$$
\Uu Q\Uu^*\;=\;-\theta^{1/2}X_1+A_1\mbox{ , }
\qquad
\Uu P\Uu^*\;=\;-\theta^{-1/2}X_2+A_2\mbox{ , }
$$

\noindent where $\langle \lL|A_1|\mM\rangle=\langle \psi_S|
\Wwth(\lL-\mM)|Q\psi_S\rangle$ and
$\langle \lL|A_2|\mM\rangle=\langle \psi_S|
\Wwth(\lL-\mM)|P\psi_S\rangle$. Because $\psi_S$, $Q\psi_S$ and
$P\psi_S$ are in $\Ss(\RM)$, $A_1$ and $A_2$ are
bounded operators. Using the standard operator inequalities
$|AB|\leq\|A\|\,|B|$ and $|A+B|\leq 2(|A|+|B|)$ and the commutation
relation $[X_1,X_2]=0$, it is now possible to deduce
$M_W(H,\Delta;q,t)\leq c_1 M_\twod(H,\Delta;q,t)+c_2$ for two positive
constants $c_1$ and $c_2$. An inequality
$M_\twod(H,\Delta;q,t)\leq c_1 M_W(H,\Delta;q,t)+c_2$ is obtained similarly.
This implies
Proposition~\ref{prop-betamagn}.

\subsection{Proof of Proposition~\ref{prop-beta2gamma} }
\label{ssec-beta2gamma}

\begin{lemma}
\label{lem-minkowskyoscill}
Let $Y_1,\ldots, Y_N$ be selfadjoint operators on $\Hh$ with common
domain which satisfy
$[Y_m,Y_n]=\imath\, c_{m,n}\1$. Then, if $c= \max_{m,n}{(|c_{m,n}|)}>0$
and if $0\leq \alpha \leq 1$, 
\begin{equation}
\label{eq-minkow_ope}
 \frac{1}{N}\,
  \sum_{n=1}^N \,
   Y_n^{2\alpha}\;
    \leq \;
 \left(
   \sum_{n=1}^N \,
    Y_n^2
  \right)^{\alpha}\;\leq\;
   \sum_{n=1}^N \,
    Y_n^{2\alpha}\;
     +\; 2N(N-1)c^{\alpha}\,.
\end{equation}
\end{lemma}

\noindent {\bf Proof: } For $\alpha =0,1$ both inequalities are trivial.
For $0<\alpha <1$ the following identity holds

\begin{equation}
\label{eq-aalpha}
A^{\alpha}\;=\;
  \frac{\sin{(\pi\alpha})}{\pi}\,
   \int_0^{\infty}\;\frac{dv}{v^{1-\alpha}}\,
    \frac{A}{v+A}\;,
\end{equation}

\noindent  for a positive operator $A$. If $A=\sum_{n=1}^N Y_n^2$
then the left-hand inequality in~(\ref{eq-minkow_ope}) follows from
$Y_n^2 \leq A$ and
from the operator monotonicity of $A/(v+A) = \1 -v/(v+A)$. On
the other hand
$$\frac{A}{v+A}\;=\;
   \sum_{n=1}^N \,
    \left(
    Y_n \frac{1}{v+A} Y_n \,+\,
     Y_n \left[Y_n, \frac{1}{v+A}\right]
    \right)\,.
$$

\noindent The first term of each summand 
is bounded by $Y_n^2/(v+Y_n^2)$.
Noting  $Y_n\left[Y_n,(v+A)^{-1}\right]=
Y_n(v+A)^{-1}\left[A,Y_n\right](v+A)^{-1}$, and
using the commutation rules for the $Y_n$'s, the second term in the r.h.s.
is estimated by
$$\left\|-2\imath\sum_{m,n} \,
    c_{m,n} \,
     Y_n\frac{1}{v+A} Y_m\frac{1}{v+A}\right\| \;\leq\;
      2 \,\frac{1}{v+c_0}\,
       \sum_{m,n} \, |c_{m,n}| \,,
$$

\noindent where $c_0$ is the infimum of the spectrum of $A$. In the
latter inequality $Y_n^2 \leq A$ has been used. By definition, there
are $m,n$ such that $c_{m,n}=c>0$ so that $Y_n^2 + Y_m^2= (Y_m-\imath
Y_n)(Y_m+\imath Y_n)+c\1 \geq c\1$. Hence $c_0 \geq c$. Integrating
over $v$, using the eq.~(\ref{eq-aalpha}), and remarking that $\sum_{m,n}
|c_{m,n}|\leq N(N-1)c\,$ gives the result.\hfill $\Box$

\vspace{.2cm}

If $S\in SL(2,\ZM)$ is a symmetry such that $S^r=1$, the 
operators $Y_n = \Ff_S^n Q \Ff_S^{-n}$ satisfy the hypothesis of 
Lemma~\ref{lem-minkowskyoscill}, because  calculating the
derivative of (\ref{eq-weylimp}) at $\aG=0$ shows that each $Y_n$ is
linear in $P$ and $Q$. Clearly
$\HGS = 1/(2r)\,\sum_{n=1}^r Y_n^2$.
If $H\in\Ath$ is $S$-invariant, then $\HGS(t) = 1/(2r)\,\sum_{n=1}^r 
\Ff_S^n Q^2(t) \Ff_S^{-n}$,
where $A(t)= e^{\imath t\,H_W} A e^{-\imath t\,H_W}$ whenever $A$ is
an operator on $\Hh$. Therefore, if $0\leq q\leq 2$, the
inequality~(\ref{eq-minkow_ope}) leads to (with
$\chi_{\Delta}=\chi_{\Delta}(H_W)$)
$$\langle \fis |\,
   \chi_{\Delta} \HGS (t)^{q/2}\,\chi_{\Delta}|
    \fis \rangle \; \leq \;
   r(2r)^{-q/2}\, 
    \langle \fis |\,
     \chi_{\Delta} |Q(t)|^{q}\chi_{\Delta} \,|
      \fis \rangle \,+\,
     2r(r-1)\left( \frac{c}{2r}\right)^{q/2}\,,
$$

\noindent where $\Ff_s\fis =\fis$ has been used. 
Proposition~\ref{prop-beta2gamma} is then a direct consequence of the
definitions of the exponents $\beta_\oned^{\pm}(H,\Delta;q),\, 
\beta_W^{\pm}(H,\Delta;q)$
and of the following lemma:

\begin{lemma}
\label{lem-qu2ex}
Let $\phi\in\Ss(\RM)$, $\theta \geq 2\pi$ and $q\geq 0$. Then, there are
two positive constants $c_0,\,c_1$ such that, for
any element $B\in\Ath$
$$\langle \phi|B_W^{\ast}\,|Q|^{q}\, B_W\,| \phi\rangle\; \leq \;
   c_0 \;
    \int_0^{2\pi}\;
     \frac{d\omega}{2\pi}\;
      \langle 0|
       B_{\omega}^{\ast}\,|\hX|^{q}\, B_{\omega}\,
        | 0\rangle \,+\,
       c_1\,,
$$

\noindent where $B_W =\piw(B)$ and $B_{\omega} =\pi_{\omega}(B)$.
\end{lemma}

\noindent {\bf Proof: } Definition~(\ref{eq-geeomega}) and 
identity~(\ref{eq-intdirrep}) of Section~\ref{ssec-rotationalg} lead to
$$\langle \phi|B_W^{\ast}\,|Q|^{q}\, B_W\,| \phi\rangle\; = \;
   \theta^{(q-1)/2}\, \int_{0}^{\theta}\,
    d\omega\,
     \sum_{n,n'\in\ZM}\,
      \overline{\phi \left(\frac{\omega -n\theta}{\sqrt{\theta}}\right)}\;
       \phi \left(\frac{\omega -n'\theta}{\sqrt{\theta}}\right)\,
       \langle n|K_{\omega}|n'\rangle\,,
$$

\noindent with $K_{\omega} = B_{\omega}^{\ast} |(\omega/\theta) -\hX|^{q}
B_{\omega}$. Since $K_{\omega}$ is a positive operator, the Schwarz
inequality gives $|\langle n|K_{\omega}|n'\rangle | \leq (
\langle n| K_{\omega} |n \rangle + \langle n'| K_{\omega}|n' \rangle)/2$.
Both terms can be bounded similarly. The covariance property of
$\pi_{\omega}$ (see Section~\ref{ssec-rotationalg}) gives
$\langle n| K_{\omega} |n \rangle = \langle 0| K_{\omega-n\theta} |0 \rangle$.
Since $\phi\in\Ss(\RM)$, summing up over $n'$ first, then over $n$,
there are constants $C,\,c_1$ such that
$$\langle \phi|B_W^{\ast}\,|Q|^{q}\, B_W\,| \phi\rangle\, \leq \,
   C \int_{\RM} \, dx \, |\phi(x)|\,
    \langle 0|\, K_{x\sqrt{\theta}}\, |0 \rangle \,\leq\,
     C \int_{\RM} \, dx \, |\phi(x)|\,
    \langle 0|\, 
     B_{x\sqrt{\theta}}^{\ast} |\hX|^{q} B_{x\sqrt{\theta}}\,
      |0 \rangle \,+\, c_1\,,
$$

\noindent where the inequality $|x-\hX|^{q} \leq C_q (|x|^{q} + |\hX|^{q})$,
valid for $q\geq 0$ and some suitable constant $C_q$, has been used.
Thanks to the periodicity of $\pi_{\omega}$, the r.h.s. of the
latter estimate can be written as
$$\mbox{\rm r.h.s.} \, \leq \,
       \int_0^{2\pi}\, \frac{d\omega}{\sqrt{\theta}}\,
        \langle 0|\, B_{\omega}^{\ast} |\hX|^{q} B_{\omega}\,|0 \rangle \,
        \sup_{0<\omega <2\pi}\,
         \,
          \sum_n
          \left|\phi \left(\frac{\omega -2\pi n}{\sqrt{\theta}}\right)\right|
\,+c_1\;,
$$

\noindent completing to the proof of the lemma. \hfill $\Box$

\vspace{.2cm}

\section{Bounds on phase-space transport}
\label{sec-proofs}

Section \ref{ssec-theorem1} is devoted to the proof of
Theorem~\ref{th-gamma2rho} 
assuming  Propositions~\ref{prop-hermiteW} and \ref{prop-mehler} which
in turn 
are proven in the subsequent sections.  

\subsection{Proof of Theorem~\ref{th-gamma2rho}}
\label{ssec-theorem1}

\noindent 
The proof goes along the lines of \cite{GSB3} and is reproduced here for 
the sake of completeness. As shown in \cite{GSB3}, the time average
$\langle f(\cdot)\rangle_T$ of a non-negative function can be replaced
by the gaussian average
$$\left\langle f(\cdot)\right\rangle_T^g \;=\;
   \int_{{\bf R}}
    \frac{dt}{2T\sqrt{\pi} }\,
    e^{-t^2/4T^2}f(t)\,,
$$

\noindent without changing the values of  the growth exponents, provided
$f$ has at most powerlaw increase. 
Let $\Delta \subset \RM$ be a Borel set and
$\psi_{\Delta} (t) = e^{-\imath t\,H_W}\chi_{\Delta}(H_W )\fis$.
Since $x^{\alpha}\geq (1-e^{-x})$ whenever $0\leq \alpha \leq 1$ and
$x\geq 0$, for any $\delta>0$ one has
$$
\left\langle M_q (H,\Delta;T)\right\rangle_T^g
\; \geq \; 
   \delta^{-q/2} 
    \left(
     \| \psi_{\Delta} \|^2\,-\,
     \left\langle \langle \psi_{\Delta}(t)|
       e^{-\delta\,\HGS}\,|
        \psi_{\Delta}(t)\rangle \right\rangle_T^g
    \right)\,.
$$

\noindent For $\Delta_1 \subset \Delta$, $\Delta_1^c$ will denote
the complement $\Delta\setminus \Delta_1$.
The decomposition of $\psi_{\Delta}$ into $\psi_{\Delta_1} +
\psi_{\Delta_1^c}$ gives rise to the following lower bound
$$
\left\langle M_q (H,\Delta;T)\right\rangle_T^g
\; \geq \; 
   \delta^{-q/2}
    \left(
     \| \psi_{\Delta_1} \|^2\,- \,
     A_{\Delta_1,\Delta_1}(T,\delta) \,-\,
      2\,\Re e\, A_{\Delta_1,\Delta_1^c}(T,\delta)
    \right)\,,
$$

\noindent where $A_{\Delta_1,\Delta_2}(T,\delta):= \left\langle \langle
\psi_{\Delta_1}(t)| e^{-\delta\,\HGS}\,|\psi_{\Delta_2}(t)\rangle
\right\rangle_T^g$. Using the spectral decomposition of $\HGS$
(see eq.~(\ref{eq-oscillatorS}) in Section~\ref{ssec-symmetries}),
it is easy to get
$$A_{\Delta_1,\Delta_2}(T,\delta) \; = \;
   \int_{\Delta_1}\,d\rhs (E)\,
    \int_{\Delta_2}\,d\rhs (E')\,
     e^{-(E-E')^2 T^2} \,
      \sum_{n=0}^{\infty} \,
       \Phi_{n,S}(E) \, \overline{\Phi_{n,S}(E')}\,
        e^{-\delta\mu (n+1/2)}\,.
$$

\noindent The Schwarz inequality
$2\,|\langle \psi_1|\psi_2\rangle| \leq \|\psi_1\|^2 + \|\psi_2\|^2\,$
applied to the sum on the r.h.s., 
together with Proposition~\ref{prop-hermiteW}, lead to
$$|A_{\Delta_1,\Delta_2}(T,\delta)| \;\leq \; 
   c_{\epsilon} \,\delta^{-(1/2 +\epsilon)} \,
    \int_{\Delta_1}\,d\rhs (E)\,
     \int_{\Delta_2}\,d\rhs (E')\,
      e^{-(E-E')^2 T^2} \,,
$$

\noindent for a suitable constant $c_{\epsilon}$. For $\alpha >0$,
let $\Delta_1=\Delta(\alpha,T)$ be chosen as
$$\Delta(\alpha,T) \;=\; 
   \left\{ E\in \Delta\;\left|\;
    T^{-\alpha-1/\log(T)} \,\leq \,
     \int_{\Delta}\,d\rhs (E')\,
      e^{-(E-E')^2 T^2} \,
      \leq T^{-\alpha} \right\}\right. \,.
$$

\noindent By definition of $\rhs$ it follows then that
$$
\left\langle M_q (H,\Delta;T)\right\rangle_T^g
\; \geq \; 
   \delta^{-q/2} \,
    \rhs(\Delta(\alpha,T))\,
     \left(
      1- c_{\epsilon} \delta^{-(1/2+\epsilon)} T^{-\alpha}
     \right)\; \geq \;
      c\, T^{q\alpha/(1+2\epsilon)} \;
       \rhs(\Delta(\alpha,T)) \,,
$$

\noindent for suitable $c_{\epsilon},\,c$, and the choice 
$\delta = (2cT^{-\alpha})^{2/(1+2\epsilon)}$.
The final step uses  Lemma~\ref{lem-BGT} below, which is a variation of a result  in \cite{BGT}. Choosing $p = 1-q/
(1+2\epsilon)$ therein, the definition of the multifractal dimensions completes 
the proof of Theorem~\ref{th-gamma2rho}.\hfill $\Box$

\begin{lemma}
\label{lem-BGT}
Let $\rho$ be a positive measure on $\RM$ with compact support $I$ and 
define for $T>0$
$$I_\alpha(T) \;=\;
   \left\{ E\in I\,\left|\,
    T^{-\alpha -1/\log(T)} \,\leq\,
     \int_{I}\,d\rho (E')\,
      e^{-(E-E')^2 T^2} \,=\,\rho(B^g_T(E)) \,\leq\,
      T^{-\alpha}\,\right.\right\}\,.
$$

\noindent Then, for all $p\in [0,1]$, there is $\alpha=\alpha(p,T)$
and a constant $c$ such that
$$
\rho(I_\alpha(T)) \;\geq\;
 \frac{c\,T^{(p-1)\alpha}}{\log(T)}\;
  \int_I d\rho(E)\; \left(\rho(B^g_T(E)) \right)^{p-1}\,.
$$
\end{lemma}

\noindent {\bf Proof: } Let $\kappa>0$ and set $\Omega_0=
\left\{ E\in \mbox{supp}(\rho) |\, \rho(B^g_T(E))\leq
T^{-\kappa} \right\}$. In addition,  for $j=1,\ldots,\kappa\log(T)$
let $\Omega_j = \left\{ E\in
\mbox{supp}(\rho)\, |\, T^{-\kappa+(j-1)/\log(T)}\leq \rho(B^g_T(E))
\leq T^{-\kappa+j/\log(T)} \right\}$. Then

\begin{equation}
\label{eq-ineq} 
\int d\rho(E)\,\rho(B^g_T(E))^{p-1} 
 \;\leq\; 
  \int_{\Omega_0}
   d\rho(E)\,
    \rho(B^g_T(E))^{p-1} +
     \kappa\log(T)\max_{j=1\ldots\kappa\log(T)}
      \int_{\Omega_j} d\rho(E)\,
       \rho(B^g_T(E))^{p-1}
\end{equation}

\noindent Let $j=j(T,p)$ be the index where the maximum is taken, and then
set $\alpha=\alpha(T,p)=\kappa-j\log(T)$. It only remains to show that the $
\Omega_0$-term is subdominant if only $\kappa$ is chosen sufficiently big.
To do so, the support of $\rho$ is covered with intervals $(A_k)_{k=1\ldots K}$
of length $1/T$. Then $K\leq T\,|\mbox{supp}(\rho)|$ (where $|A|$ denotes the 
diameter of $A$). If $a_k=\inf\{\rho(B^g_T(E))|E\in A_k\cap \Omega_0\}$,
then $a_k\leq T^{-\kappa}$ by definition of $\Omega_0$. 
Moreover $\rho(B^g_T(E)) \geq \int_{A_k\cap\Omega_0} d\rho(E') e^{-(E-E')^2T^2}$.
In particular, if $E\in A_k\cap\Omega_0$, then $|E-E'| T \leq 1$ implying
$\rho(B^g_T(E)) \geq e^{-1} \rho(A_k\cap\Omega_0)$ and thus,
 $\rho(A_k\cap\Omega_0)\leq e a_k$. Hence ($p-1\leq 0$):

$$\int_{\Omega_0} d\rho(E)\;
   \rho(B^g_T(E))^{p-1} \;\leq\;
    \sum_{k\leq K}
     \rho(A_k\cap\Omega_0) a_k^{p-1} \;\leq\;
      e\sum_{k\leq K}a_k^p \;\leq\;
       e \, T^{1-\kappa p}\,|\mbox{supp}(\rho)| \mbox{ . } 
$$

\noindent Hence choosing $\kappa=2/p$, for example, provides a subdominant
contribution in (\ref{eq-ineq}) such that (\ref{eq-ineq}) fulfills the desired
bound. \hfill $\Box$

\subsection{Proof of Proposition~\ref{prop-hermiteW}}
\label{ssec-proposition3}

This section is devoted to the proof of Proposition~\ref{prop-hermiteW}
assuming Proposition~\ref{prop-mehler}. Since $\hU =e^{2\imath \pi Q/\sqrt{\theta}} =
\Ww_{\theta'}(0,1)$ commutes with $\piw (\Ath)$, it commutes, in particular,
with $H_W$. Therefore the pair $(H_W,\hU)$ has a joint spectrum contained
in $\RM\times \TM$. Let $m_S$ denote the spectral measure of the pair
relative to $\fis$ defined by
$$\int_{\RM\times \TM}\, dm_S(E,\eta)\, F(E,e^{\imath \eta})\;=\;
   \langle \fis |\, F(H_W,\hU)\,| \fis \rangle\,,
\hspace{2cm}
  \forall\, F\in\Cc_0(\RM\times \TM)\,.
$$

\noindent The marginal probabilities associated with $m_S$ are respectively
$d\rhs (E)$, the spectral measure of $H_W$, and 
$d\eta\, \|\Gg_{\theta\,\eta/2\pi}
\fis\|_{\ell^2}^2 \,\theta/(2\pi)$ 
for $\eta\in\TM$, the spectral measure of $\hU$. 
Thanks to the Radon-Nikodym theorem, $m_S$ can be written either as
\begin{equation}
\label{eq-condexpeta}
\int_{\RM\times \TM} dm_S(E,\eta)\; F(E,e^{\imath \eta})\; =\;
   \frac{\theta}{2\pi}
    \int_0^{2\pi} d\eta  
     \int_{\RM}\, d\mu_{(\theta\eta/2\pi)} (E)\; F(E,e^{\imath \eta})\,,
\end{equation}

\noindent (where $\mu_{\omega}$ is the spectral measure of $H_{\omega}$ relative
to $\Gg_{\omega}\fis$), or as 
\begin{equation}
\label{eq-condexpE}
\int_{\RM\times \TM} dm_S(E,\eta) \;F(E,e^{\imath \eta})\; =\; 
   \int_{\RM} d\rhs (E)
    \int_0^{2\pi} d\nu_E(\eta)\; F(E,e^{\imath \eta}) \,,
\end{equation}

\noindent for some probabilty measure $\nu_E$ depending $\rhs$-measurably 
upon $E$. Due to the spectral theorem, for every $n\in\ZM$, there is 
a function $g_n(\omega,\cdot) \in L^2(\RM, \mu_{\omega})$ such that
\begin{equation}
\label{eq-cyclicforn}
\langle \Gg_{\omega} \fis |\, f(H_{\omega})\, |n \rangle\; = \;
   \int_{\RM} d\mu_{\omega}(E) \;
    f(E)\, g_n(\omega,E)\,.
\end{equation}

\noindent In the following lemma, 
$\tilde{g}_n (\eta,E)$ stands for $\theta^{-1/4} 
g_n(\theta \eta/2\pi, E)$:

\begin{lemma}
\label{lem-cyclicfis}
Let $\psi\in \Ss (\RM)$. Then the representative in $L^2(\RM,\rhs)$ of the 
projection of $\psi$ on the $H_W$-cyclic component of $\fis$ is given by
$$\tilde{\psi}(E)\;=\;
   \int_0^{2\pi} d\nu_E(\eta)\;
    \sum_{n\in\ZM}\,\tilde{g}_n (\eta,E)\;
     \psi\left((\eta -2\pi n)\,\theta^{1/2}/2\pi
         \right)
$$
\end{lemma}

\noindent {\bf Proof: } $\tilde{\psi}$ is defined by $\langle \fis |\,
f(H_W)\, |\psi \rangle = \int_{\RM} d\rhs(E) f(E) \tilde{\psi}(E)$ for every 
$f\in\Cc_0(\RM)$. On the other hand, thanks to eq.~(\ref{eq-intdirrep}),

$$\langle \fis |\, f(H_{\omega})\,| \psi \rangle \;=\;
   \int_0^{\theta} d\omega \;
    \langle \Gg_{\omega}\fis |f(H_{\omega})| \Gg_{\omega}\psi \rangle \;=\;
     \sum_{n\in\ZM} 
     \int_0^{\theta} d\omega \;
      \langle \Gg_{\omega}\fis |f(H_{\omega}) \,|n\rangle \;
       (\Gg_{\omega}\psi)(n)\,.
$$

\noindent Then, using the definition (\ref{eq-cyclicforn}) of $g_n$ together 
with eqs.~(\ref{eq-condexpeta}) and 
(\ref{eq-condexpE}), and changing from $\omega$ 
to $\eta$, gives the result.\hfill $\Box$

\vspace{.2cm}

\noindent {\bf Proof of Prop.~\ref{prop-hermiteW}:} Let $\Delta \subset \RM$
be a Borel set and, for $\delta >0$, let $Q(\Delta,\delta)$ be defined by
$$Q(\Delta,\delta)\;=\;
   \int_{\Delta} d\rhs (E)\;
    \sum_{n=0}^{\infty}\,
     e^{-\delta (n+1/2)}\, 
      |\Phi_{n,S}(E)|^2\,.
$$

\noindent Thanks to Lemma~\ref{lem-cyclicfis} applied to the eigenstates
$\phi_S^{(n)}$ of $\HGS$ (see eq.~(\ref{eq-oscillatorS})), it can be 
written as
$$
\begin{array}{lcl}
Q(\Delta,\delta) & = &
   \int_{\Delta} d\rhs \int d\nu_E(\eta) d\nu_E(\eta')\;
    \sum_{m,m'} \tilde{g}_m (\eta,E)\,
     \overline{\tilde{g}_{m'} (\eta,E)} \cdots \\
    &  &  \\
 & & \cdots   \sum_{n=0}^{\infty}\,
      e^{-\delta (n+1/2)}\;
       \phi_S^{(n)}((\eta -2\pi m)\,\theta^{1/2}/2\pi)\,
        \overline{\phi_S^{(n)}((\eta' -2\pi m')\,\theta^{1/2}/2\pi)}\,.
\end{array}
$$
The last sum on the r.h.s. of this identity reconstructs 
the Mehler kernel of eq.~(\ref{eq-mehlerS}) with $t=\delta/\mu$. 
It will be convenient to define
\begin{equation}
\label{eq-gedelta}
G_{\delta}(E;x) \;=\;
   \int d\nu_E(\eta')\sum_{m'}\;
    \left| \Mm_S(\delta/\mu;x, (\eta -2\pi m')\,\theta^{1/2}/2\pi) \right|\,.
\end{equation}

\noindent Since the Mehler kernel decays fastly, this sum converges.
Using the Schwarz inequality together with the symmetry $(m,\eta)
\leftrightarrow (m',\eta')$, $Q(\Delta,\delta)$ can be bounded from above by
$$Q(\Delta,\delta) \;\leq\;
   \sum_{m}\,
    \int_{\Delta} d\rhs \int d\nu_E(\eta)\;
     \left|\tilde{g}_m (\eta,E)\right|^2\;
      G_{\delta}\left(E;(\eta -2\pi m)\,\theta^{1/2}/2\pi \right)\,.
$$

\noindent Thanks to eqs.~(\ref{eq-condexpeta}) 
and (\ref{eq-condexpE}), and changing
again from $\eta$ to $\omega$, this bound can be written as
$$Q(\Delta,\delta) \;\leq\;
   \sum_{m}\,
    \int_0^{\theta} \frac{d\omega}{\theta^{1/2}} 
     \int_{\Delta} d\mu_{\omega}(E)\;
     \left| g_m (\omega,E) \right|^2\;
      G_{\delta}\left(E;(\omega -m\theta)/\theta^{1/2} \right)\,.
$$

\noindent If now $P_{\omega}$ is the projection on the $H_{\omega}$-cyclic 
component of $\Gg_{\omega}\fis$ in $\ell^2(\ZM)$, the 
definition~(\ref{eq-cyclicforn}) of $g_m$ and the covariance lead to the
following inequality
$$\int d\mu_{\omega}(E)\; 
   \left| g_m (\omega,E) \right|^2\,f(E) \;=\;
    \langle m|P_{\omega} f(H_{\omega}) P_{\omega}| m\rangle \;\leq\;
     \langle 0| f(H_{\omega-m\theta}) | 0\rangle \,,
$$

\noindent valid for $f\in\Cc_0(\RM),\; f\geq 0$, because $H_{\omega}$ 
commutes with $P_{\omega}$ and the latter is a projection. Let then 
$\mu_{\omega}^{(0)}$ be the spectral measure of $H_{\omega}$  relative to the
vector $|0\rangle$. The previous estimate implies
$$\begin{array}{lcl}
Q(\Delta,\delta) & \leq &
   \sum_{m}\, \theta^{-1/2}
    \int_0^{\theta} d\omega 
     \int_{\Delta} d\mu_{\omega-m\theta}^{(0)}(E)\;
      G_{\delta}\left(E;(\omega -m\theta)/\theta^{1/2} \right) \\
    &  &  \\
 & \leq  & \theta^{-1/2} 
    \int_{\infty}^{\infty} d\omega
     \int_{\Delta} d\mu_{\omega}^{(0)}(E)\;
      G_{\delta}\left(E;\omega/\theta^{1/2} \right)\,.
\end{array}
$$

\noindent Since $\mu_{\omega}^{(0)}$ is $2\pi$-periodic with respect
to $\omega$, 
the latter integral can be decomposed into a sum over 
intervals of length $2\pi$
leading to the following estimate
$$Q(\Delta,\delta) \; \leq \;
   \theta^{-1/2}
    \int_0^{2\pi} d\omega
     \int_{\Delta} d\mu_{\omega}^{(0)}(E)\;
      \sum_{k\in\ZM}\;
       G_{\delta}\left(E;(\omega+2\pi k)/\theta^{1/2} \right)\,.
$$

\noindent Definitions~(\ref{eq-trace1D}) of the trace on $\Ath$, 
(\ref{eq-DOSdef}) of the DOS and (\ref{eq-gedelta}) of $G_{\delta}$ give

$$Q(\Delta,\delta) \; \leq \;
   \frac{2\pi}{\theta^{1/2}}
    \int_{\Delta\times [0,2\pi]}
     d\Nn (E)\,d\nu_E(\eta) 
      \sum_{(k,m)\in\ZM^2}\;
    \left| 
     \Mm_S \left(
 \delta/\mu; \frac{\omega+2\pi k}{\theta^{1/2}},\,
    \frac{(\eta -2\pi m)\,\theta^{1/2}}{2\pi}
      \right) \right|\,.
$$

\noindent The result of  Proposition~\ref{prop-mehler} can now be used. 
Remarking that $\nu_E$ is a probability, and using the equivalence 
between $\rhs$ and the DOS (Theorems~\ref{theo-grounstate}
and \ref{theo-rhophivsNn} combined), the last 
estimate implies
$$Q(\Delta,\delta) \; \leq \;
   c_{\epsilon}\, \rhs(\Delta)\; \delta^{-(1/2+\epsilon)}\,,
$$

\noindent for some suitable constant $c_{\epsilon}$. Since this inequality holds for 
all Borel subset $\Delta$ of $\RM$, the Proposition~\ref{prop-hermiteW} 
is proven. \hfill $\Box$

\subsection{Proof of Proposition~\ref{prop-mehler}}
\label{ssec-proposition4}

\noindent If $\alpha =\theta/2\pi \in [0,1]$ is an irrational number, 
a rational approximant 
is a rational number $p/q$, with $p,q$ prime to each other, such that 
$|\alpha-p/q|< q^{-2}$. The continued fraction expansion
$[a_1,\cdots,a_n,\cdots ]$ of $\alpha$ \cite{Her},
provides an infinite sequence $p_n/q_n$ of such approximants, the 
{\em principal convergents},  recursively defined by $p_{-1} =1 ,q_{-1} = 0, 
p_0 =0, q_0 = 1$ and $s_{n+1}=a_{n+1} s_{n}+s_{n-1}$ if $s=p,q$.
It can be proved (see \cite{Her} Prop. 7.8.3) that $\alpha$ is a
number of Roth type  (see eq.~(\ref{eq-roth}) in Section \ref{sec-notresults})
if and only if $\sum_{n=1}^{\infty} a_{n+1}/q_n^{\epsilon} <\infty$ for 
all $\epsilon >0$.

\vspace{.2cm}

The proof of Proposition~\ref{prop-mehler} relies upon
the so-called {\em Denjoy-Koksma inequality} \cite{Her}. Let $\varphi$ 
be a periodic function on $\RM$ with period $1$, of bounded 
total variation $\Var(\varphi)$ over a period interval.  Then (see
\cite{Her}, Theo. 3.1)

\vspace{.3cm}

\noindent {\bf Theorem [Denjoy-Koksma inequality]}
{\it Let $\alpha\in [0,1]$ be irrational and let $\varphi$  be a real
valued function on $\RM$ of period one. Then, if $p/q$ is a rational
approximant of $\alpha$
$$\left|
   \sum_{j=1}^q \varphi(x+j\alpha)\,
    - q\int_0^1 dy\, \varphi (y) \,
   \right| \;\leq\; \Var(\varphi)\,.
$$
}

\vspace{.3cm}

Proposition~\ref{prop-mehler} is a direct consequence
of the definition of the Mehler kernel (see eq.~(\ref{eq-mehlerS})) and of the
following result

\begin{lemma}
\label{lem-rothmehler}
If $\delta >0$, let $F_{\delta}$ be the function on $\RM^2$ defined by
$$F_{\delta}(x,y)\;=\; 
   \delta\, (x+y)^2 + \delta^{-1} \, (x-y)^2 \,.
$$

\noindent If $\alpha$ is a number of Roth type, then for any
$a>0,\,\epsilon >0$, 
there is $c_\epsilon >0$ such that
$$\sup_{x,y \in \RM}
   \sum_{(k,m)\in\ZM^2}
    e^{-a\,F_{\delta}(x+k,y+m\alpha)} \;\leq\;
     c_\epsilon\, \delta^{-\epsilon}\,,
\hspace{1cm}
 \forall\;\; \delta \in (0,1)\,.     
$$
\end{lemma}

\noindent {\bf Proof: } Let $(x_0,y_0)\in \RM^2$ be fixed and set $\Ll=
\{ (x_0+k,y_0+m\alpha) \in\RM^2 \,|\, (k,m)\in\ZM^2 \}$. If $S(x_0,y_0)= 
\sum_{k,m} e^{-a\,F_{\delta}(x_0+k,y_0+m\alpha)}$ then $S$ is periodic of
period $1$ in $x_0$ and of period $\alpha $ in $y_0$. Therefore, it is enough 
to assume $0\leq x_0<1$ and $0\leq y_0 <1$ (since $0 <\alpha <1$). For 
$0<\sigma <1$ and for $j\in\NM$, 
let $\Ll_j$ be the set of points $(x,y)\in\Ll$ 
for which $j^2 \delta^{-\sigma} \leq F_{\delta}(x,y) <(j+1)^2
\delta^{-\sigma}$. Thus
\begin{equation}
\label{eq-mehlerdiscrete}
S(x_0,y_0) \; \leq \;
 \sum_{j=0}^{\infty}
  e^{-aj^2\delta^{-\sigma}}\;
   |\Ll_j|\;,
\end{equation}

\noindent where $|A|$ denotes the number of points in $A$.
$\Ll_j$ is contained in an elliptic crown with axis along the two diagonals
$x=\pm y$. In particular,
\begin{equation}
\label{eq-conditiononLl}
(x,y)\in\Ll_j\;\; \Rightarrow \;\;
   \max\{|x|,|y|\}\,\leq\, (j+1) \, \delta^{-(1+\sigma)/2}\;\;\;
    \mbox{\rm and} \;\;\;
     |x-y|\,\leq\, (j+1) \, \delta^{(1-\sigma)/2}\,.
\end{equation}

\noindent If $j\geq 1$, the number of points contained in $\Ll_j$ can be 
estimated by counting the number of rectangular cells of sizes $(1,\alpha)$
centered at points of $\Ll$ and meeting the elliptic crown. Since this crown
is included inside the square $\max\{|x|,|y|\} \leq (j+1)\delta^{-(1+\sigma)/2}$
it is enough to count such cells meeting this square. Such cells are all included
inside the square $C=\{ (x,y)\in\RM^2\,|\, \max\{|x|,|y|\}\leq
(j+2) \delta^{-(1+\sigma)/2}\}$ (since $\delta \leq 1$). Hence the number of such
cells is certainly dominated by the ratio of the area of $C$ to the area of each
cell, namely 
$$|\Ll_j| \;\leq \;
    \frac{(j+2)}{\alpha}^2 \;\delta^{-(1+\sigma)}\,.
$$

\noindent Therefore, the part of the sum in~(\ref{eq-mehlerdiscrete}) coming from
$j\geq 1$ converges to zero as $\delta\downarrow 0$. In particular, it is 
bounded by a constant $c_1$ that is independent of $(x_0,y_0)$. 
Thus, it is sufficient to consider the term $j=0$ only.

\vspace{.2cm}

Let $\varphi$ be the function on $\RM$ defined by $\varphi (x)
\;=\; \sum_{k\in\ZM} \chi_I (x+y_0-x_0 +k)$ where $I$ is the interval
$I=[-\delta^{(1-\sigma)/2},\delta^{(1-\sigma)/2}]\subset \RM$. 
It is a periodic function of period $1$ with $\Var (\varphi)=2$. Moreover,
using (\ref{eq-conditiononLl}) it can be checked easily that
$$S(x_0,y_0) \; \leq \;
   c_1 +
    \sum_{|m|< M}
     \varphi (m\alpha) \; \leq \;
      c_1 + \sum_{m=0}^{M-1}
       \left(\varphi (m\alpha)+\varphi (-m\alpha)\right)\,,
$$

\noindent provided 
$M\geq 3\,\delta^{-(1+\sigma)/2}/\alpha$. For indeed, $(x,y)\in
\Ll_0$ only if $|y_0+m\alpha|\leq \delta^{-(1+\sigma)/2}$ for some
$m\in\ZM$.  Let then $n\in\NM$ 
be such that $q_n \leq M <q_{n+1}$, where the $p_n/q_n$'s are the principal 
convergents of $\alpha$. Replacing $M$ by $q_{n+1}$ in the r.h.s. 
gives an upper bound. By the Denjoy-Koksma inequality, the r.h.s. is
therefore bounded from above by $c_1 + 4q_{n+1}\delta^{(1+\sigma)/2}$. 
Since $\alpha$ is a number of Roth type, $q_{n+1} \leq (a_{n+1}+1)q_n \leq 
c_2\cdot q_n^{1+\sigma}$, 
thanks to Prop. 7.8.3 in \cite{Her} (see above). It is important to notice that
$c_2$ only depends upon $\alpha$ and the choice of the exponent $\sigma$. Collecting
all inequalities, gives
$$S(x_0,y_0) \; \leq \;
   c_1 + \frac{12\cdot c_2}{\alpha}\; \delta^{-2\sigma}\,.
$$

\noindent Choosing $\sigma = \epsilon/2$ and remarking that none of the constants
on the r.h.s. depends on $(x_0,y_0)$ leads to the result. \hfill $\Box$


\vspace{.2cm}

\section*{Appendix: Proofs of various results on Weyl operators}
\label{appendix}

\noindent {\bf Proof} of eqs. (\ref{eq-WeylHusimi}) and
(\ref{eq-HusimiWeyl}): Due to the polarization principle,  
(\ref{eq-WeylHusimi}) is equivalent to

\begin{equation}
\label{eq-WeylHusimi2}
\langle \phi |\WG (\aG) \,|\phi \rangle\; 
\overline{\langle \psi |\WG (\aG) \,|\psi \rangle\; }
\; =\;
\int_{\RM^2}\, \frac{d^2 \bG}{2\pi}\,
e^{\imath \,\aG\wedge \bG}\;
   \left|\langle \psi |\WG (\bG)\,|\psi\rangle\right|^2 
\mbox{ . }
\end{equation}

\noindent By inverse Fourier transform, (\ref{eq-WeylHusimi2}) is
equivalent to 

\begin{equation}
\label{eq-WeylHusimi3}
\left|\langle \phi |\WG (\bG)\,|\psi\rangle\right|^2 
\;=\;
\int_{\RM^2}\, \frac{d^2 \aG}{2\pi}\,
e^{\imath \,\bG\wedge \aG}\;
\langle \phi |\WG (\aG) \,|\phi \rangle\; 
\overline{\langle \psi |\WG (\aG) \,|\psi \rangle\; }
\mbox{ , }
\end{equation}

\noindent which is equivalent to (\ref{eq-HusimiWeyl}), so that it is
sufficient to prove (\ref{eq-WeylHusimi2}). Using
(\ref{eq-Weyldef}), 
$$
\mbox{r.h.s. of (\ref{eq-WeylHusimi2})}
\; =\;
\int_{\RM^2}
\frac{db_1\,db_2}{2\pi}
\int_\RM dx\int_\RM dy \;
\overline{\phi(x)}\, 
\phi(y)\,
\psi(x+b_1)\,
\overline{\psi(y+b_1)}
\,
e^{\imath(b_2(x-y+a_1)-a_2b_1)}
\mbox{ . }
$$

\noindent The integral over $b_2$ can be immediately evaluated by
$\int_\RM db_2 \,e^{\imath b_2(x-y+a_1)}=2\pi\,\delta(y-x-a_1)$. Thus
the integration over $y$ is elementary. Changing variable from $b_1$
to $x'=x+b_1$ therefore gives

$$
\mbox{r.h.s. of  (\ref{eq-WeylHusimi2})}
\; =\;
\int_\RM dx\int_\RM dx'
\;
\overline{\phi(x)}\, 
\phi(x+a_1)\,
e^{\imath a_2x+\imath \frac{a_1a_2}{2}}
\;
\psi(x')\,
\overline{\psi(x'+a_1)}
\,e^{-\imath a_2x'-\imath \frac{a_1a_2}{2}}
\mbox{ , }
$$

\noindent which is precisely the l.h.s. of (\ref{eq-WeylHusimi2}).
\hfill $\Box$

\vspace{.2cm}

\noindent {\bf Proof} of eq. (\ref{eq-intdirrep}):
It is sufficient to verify (\ref{eq-intdirrep}) for the generators 
$A=W_\theta(m)$, $m\in\ZM^2$, of $\Aa_\theta$. For such $A$, 

$$
\mbox{r.h.s. of (\ref{eq-intdirrep})}
\;=\;
\int^\theta_0 \frac{d\omega}{\sqrt{\theta}}
\;\sum_{n,l\in\ZM}
\,
\overline{\phi\left(\frac{\omega - n\theta}{\sqrt{\theta}}\right)}
\;
\langle n| \pi_\omega(W_\theta(m)|l\rangle\;
\psi\left(\frac{\omega - l\theta}{\sqrt{\theta}}\right)
\mbox{ . }
$$

\noindent As  $\langle n| \pi_\omega(W_\theta(m)|l\rangle= 
e^{\imath \theta m_1m_2/2}\,e^{\imath (\omega-l\theta)m_2}
\delta_{n,l+m_1}$, the sum over $n$ can be immediately computed, and
the one  over $l$ can be combined with the integral over $\omega$ in order
to give

$$
\mbox{r.h.s. of (\ref{eq-intdirrep})}
\;=\;
\int_\RM \frac{dx}{\sqrt{\theta}}\;
\overline{\phi\left(\frac{x - m_1\theta}{\sqrt{\theta}}\right)}
\;
e^{\imath \theta \frac{m_1m_2}{2}}\,e^{\imath xm_2}
\psi\left(\frac{x}{\sqrt{\theta}}\right)
\mbox{ . }
$$

\noindent Changing variable $y=(x-m_1\theta)/\sqrt{\theta}$ and
identifying $\WG(\sqrt{\theta}m)$ shows

$$
\mbox{r.h.s. of (\ref{eq-intdirrep})}
\;=\;
\int_\RM dy\; \overline{\phi(y)}\,\left(\WG(\sqrt{\theta}m)\psi\right)(y)
\mbox{ , }
$$

\noindent namely the l.h.s. of (\ref{eq-intdirrep}) 
\hfill $\Box$

\vspace{.2cm}

\noindent {\bf Proof} of Proposition \ref{prop-Poissonsum}:
For $f\in \Ss (\RM^2)$, let $\tilde{f}$ be its symplectic Fourier
transform defined by ($\lG,\mG\in\RM^2$): 

$$
\tilde{f}(\lG)
\;=\;
\int_{\RM^2}
\frac{d^2\mG}{2\pi}
\;e^{\imath \lG\wedge\mG}\,f(\mG)
\mbox{ , }
\qquad
\Leftrightarrow
\qquad
{f}(\mG)
\;=\;
\int_{\RM^2}
\frac{d^2\lG}{2\pi}
\;e^{\imath \mG\wedge\lG}\,\tilde{f}(\mG)
\mbox{ . }
$$

\noindent Then the classical Poisson summation formula reads

$$
\sum_{m\in\ZM^2}
f(m)
\;=\;
2\pi\;\sum_{l\in\ZM^2}
\tilde{f}(2\pi\,l)
\mbox{ . }
$$

\noindent Setting $f(\mG)=
\langle \phi |\,\WG (\sqrt{\theta}\mG) \,|\phi \rangle\; 
\overline{\langle \psi |\WG (\sqrt{\theta}\mG) \,|\psi \rangle}$,
equation (\ref{eq-WeylHusimi3}) leads to

$$
\tilde{f}(\lG)
\;=\;
\frac{1}{\theta}\;
\left|\langle \psi |\,\WG
\left(\frac{2\pi}{\sqrt{\theta}}\;\lG\right)\,|\phi\rangle\right|^2 
\mbox{ . }
$$

\noindent Inserting this into the Poisson summation formula and
recalling the notation (\ref{eq-Weylredef})
gives (\ref{eq-PoissonforT}).
\hfill $\Box$

\vspace{.2cm}

\noindent {\bf Proof} of eq. (\ref{eq-psipsi}): 
By (\ref{eq-AthFourier}) and (\ref{eq-Weylredef}),
$\piw(A)= \sum_{\lL\in\ZM^2}\, a_{\lL}\, \Wwth(\lL)$  
with $a_{\lL} \,=\, \TV_{\theta} (\Wth (\lL)^{-1} \,A)$.
Thus

$$
\langle \psi |\,\piw (A) \,|\psi\rangle \;=\;
\sum_{\lL\in\ZM^2}\, a_{\lL}\;
\langle \psi |\,\Wwth(\lL) \,|\psi\rangle
\;=\;
\TV_{\theta} \left(\,
\sum_{l\in\ZM^2}
\langle \psi |\,\Wwth(\lL) \,|\psi\rangle
\Wth (\lL)^{-1}   A\right)
\mbox{ . }
$$

\noindent Comparing with the Poisson summation formula
(\ref{eq-PoissonforT}) shows (\ref{eq-psipsi}).
\hfill $\Box$

\vspace{.2cm}

\noindent {\bf Proof} of Proposition \ref{prop-sl2rweyl}:
Because of the freedom of phase and relation (\ref{eq-Weylccr}), it is
sufficient to verify all implementation formulas (\ref{eq-weylimp}) for the Weyl
operators $e^{\imath Q}$ and $e^{\imath P}$ or equivalently 
(on the domain of) their generators $Q$ and $P$.
Concerning the first formula in (\ref{eq-sl2rweyl}), it thus follows
from the identities

$$
e^{-\imath\,\kappa\,Q^2/2}\,Q\,
e^{\imath\,\kappa\,Q^2/2}
\;=\;
Q
\mbox{ , }
\qquad
e^{-\imath\,\kappa\,Q^2/2}\,P\,e^{\imath\,\kappa\,Q^2/2}
\;=\;
\kappa\, Q + P
\mbox{ . }
$$

\noindent Next let us consider the dilations on $L^2(\RM)$ defined by
$(D(a)\phi)(x)=\sqrt{e^a}\,\phi(e^a\,x)$. It generators are computed
by

$$
\left.\frac{d}{da}(D(a)\phi)(x)\right|_{a=0}
\;=\;\frac{\imath}{2} (QP+PQ)\, \phi(x)
\mbox{ , }
$$

\noindent so that for $a=-\mbox{ln}(\lambda)$

$$
e^{-\imath\,\ln{\lambda}\,(QP+PQ)/2}
\,\phi(x)\;=\;
\sqrt{\frac{1}{\lambda}}
\;\phi\left(\frac{x}{\lambda}\right)
\mbox{ . }
$$

\noindent This immediately allows to verify

$$
e^{-\imath\,\ln{\lambda}\,(QP+PQ)/2}\,Q\,
e^{\imath\,\ln{\lambda}\,(QP+PQ)/2}
\;=\;\frac{1}{\lambda}\,Q
\mbox{ , }
\qquad
e^{-\imath\,\ln{\lambda}\,(QP+PQ)/2}\,P\,
e^{\imath\,\ln{\lambda}\,(QP+PQ)/2}
\;=\;\lambda\,P
\mbox{ , }
$$

\noindent which proves the second formula in (\ref{eq-sl2rweyl}).
To prove the last one, 
we use the annihiliation-creation operators
$a=(Q-\imath P)/\sqrt{2}$ and $a^*=(Q+\imath P)/\sqrt{2}$. As
$(P^2+Q^2-1)/2=a^*a$ and $e^{-\imath sa^* a}a e^{\imath s a^*
a}=e^{\imath s}a$, the formula follows after decomposing $\WG(\aG)$
into $a$ and $a^*$. Finally we search the integral kernel 
for $K=e^{-\imath s a^*a}$, notably $(K\phi)=\int dy
\,k(x,y)\phi(y)$. If
$\phi^{(n)}_{S_4}$ are the Hermite functions,
then $K\phi^{(n)}_{S_4}=e^{\imath sn}
\phi^{(n)}_{S_4}$. Equivalently, $k$ has to satisfy $a_y k = e^{\imath
s}a_x^* k$ and $K\phi^{(0)}=\phi^{(0)}$ (here the index on the $a$'s
indicate with respect to which variable the operator acts). An Ansatz
$k(x,y)=e^{-b(x^2+y^2)+cxy+d}$ leads to the integral kernel in
(\ref{eq-frotation}).
\hfill $\Box$

\vspace{.2cm}

\noindent {\bf Proof} of Proposition \ref{eq-mehlerS}:
Let us set

$$
R\;=\;\left( \begin{array}{cc}\cos{\gamma} & -\sin{\gamma}\\
\sin{\gamma} & \cos{\gamma}\end{array} \right) 
\mbox{ , }
\qquad
D\;=\;
\left( \begin{array}{cc}\lambda & 0 \\
0 & \frac{1}{\lambda}\end{array} \right)
\mbox{ . } 
$$

\noindent Then, using the notations and formulas in Subsection
\ref{ssec-symmetries}, 

\begin{equation}
\label{eq-Htrafo}
\HGS\;=\;\frac{\mu}{2} \,\langle (RD)^tK\,|(RD)^tK\rangle
\;=\;
\mu\, \Ff_R\,\Ff_D\,
{\mathfrak H}_{S_4}\,
\Ff_D^{-1}\,\Ff_R^{-1}
\mbox{ , }
\qquad
\phi_S\;=\;\Ff_R\,\Ff_D\,\phi_{S_4}
\mbox{ . }
\end{equation}

\noindent Now $\phi_{S_4}$ is known to be the normalized
gaussian. Using the implementation formulas of Proposition
\ref{prop-sl2rweyl}, it is straightforward to calculate the gaussian
integrals giving (\ref{eq-groundstate}). The Mehler kernel $\Mm_{S_4
}(t;x,y)$ 
for ${\mathfrak H}_{S_4}=(P^2+Q^2)/2$ is well-known (and can be read
of (\ref{eq-frotation}) at imaginary time). Using (\ref{eq-Htrafo})
and the definition (\ref{eq-mehler}),

$$
\Mm_S (t;x,y)\;=\;
\int_\RM dx'\int_\RM dy'\;
\langle x |\,\Ff_R\,\Ff_D \,|x'\rangle
\;\Mm_{S_4} (t;x',y')\;
   \langle y' |\,\Ff_D^{-1}\,\Ff_R^{-1}\,|y\rangle 
\mbox{ . }
$$

\noindent The gaussian integrals herein give rise to (\ref{eq-mehlerS}).
\hfill $\Box$

\vspace{.2cm}

Let us conclude with the proof of the complementary result given in
Section \ref{sec-notresults}.

\vspace{.2cm}

\noindent {\bf Proof} of Proposition
\ref{prop-multiplicity}: The commutant $\Bb$ of the abelian
C$^\ast$-algebra generated by $H_W$ contains the
commutant of $\piw(\Ath)$, that is the von Neumann algebra
$\overline{\piw(\Athp)}$ generated by 
$\piw(\Athp)$. As $\overline{\piw(\Athp)}$ is of type II$_1$ \cite{Sak}, 
there exist $\ast$-endomorphisms $\eta_q:\mbox{Mat}_{q\times q}\to\Bb$ 
for every $q\in\NM$ (here $\mbox{Mat}_{q\times q}$ denotes the complex
$q\times q$ matrices). 

According to the spectral theorem, $\Hh$ decomposes according to the
multiplicity of $\piw(H)$: $\Hh=\oplus_{n\geq 1} L^2(X_n,\mu_n)\otimes
\CM^n \oplus L^2(X_\infty,\mu_\infty)\otimes \ell^2(\NM)$ where the
$\mu_n$'s are positive measures with pairwise disjoint supports
$X_n\subset\RM$. In this representation, $\piw(H)=\oplus_{n\geq 1}
\mbox{Mult}(E) \otimes {\bf 1}_n\oplus
\mbox{Mult}(E) \otimes {\bf 1}_\infty$ (here $\mbox{Mult}(E)$
denotes the multiplication by the identity on $\RM$)
and
$\Bb=\oplus_{n\geq 1} L^\infty(X_n,\mu_n)\otimes
\mbox{Mat}_{n\times n} 
\oplus L^\infty(X_\infty,\mu_\infty)\otimes \Bb(\ell^2(\NM))$.
Let $P_n$ be the projection on
$L^2(X_n,\mu_n)\otimes\CM^n$. Then $P_n\Bb P_n=
L^\infty(X_n,\mu_n)\otimes\mbox{Mat}_{n\times n}$.
Moreover $\phi_{n,x}(B)=P_n BP_n(x)$ defines a $\ast$-endomorphism
from $\Bb$ to $\mbox{Mat}_{n\times n}$ for $\mu_n$-almost all $x\in
X_n$. Combining with $\eta_q$, one gets $\ast$-endomorphisms
$\phi_{n,x}\circ \eta_q:\mbox{Mat}_{q\times q}\to
\mbox{Mat}_{n\times n}$ for any q satisfying 
$\phi_{n,x}\circ \eta_q({\bf 1}_q)={\bf 1}_n$. This is impossible for
any $q>n$ so that $X_n=\emptyset$ for all $n\geq 1$.

If $H_W$ had a cyclic vector, its spectrum would be
simple.
\hfill $\Box$

\vspace{.3cm}


\end{document}